\documentclass[a4paper,12pt]{article}

\usepackage{amsmath}
\usepackage[final]{graphicx}
\usepackage{bm}
\usepackage{amssymb}

\usepackage{cancel}

\usepackage{amssymb}
\newcounter{comment}

{\refstepcounter{comment}%
\begin{quote}
\ttfamily\small$\blacksquare$ \textbf{\underline{Comment} $\sharp$\thecomment:}}%
{\end{quote}}

\begin{document}
\hfill
\begin{minipage}{20ex}\small
ZAGREB-ZTF-11-04\\
\end{minipage}

\begin{center}
\baselineskip=2\baselineskip
\textbf{\LARGE{
Exotic Seesaw-Motivated Heavy Leptons
 at the LHC
}}\\[6ex]
\baselineskip=0.5\baselineskip

{\large Kre\v{s}imir~Kumeri\v{c}ki,
Ivica~Picek
and
Branimir~Radov\v{c}i\'c
}\\[4ex]
\begin{flushleft}
\it
Department of Physics, Faculty of Science, University of Zagreb,
 P.O.B. 331, HR-10002 Zagreb, Croatia\\[3ex]
\end{flushleft}
\today \\[5ex]
\end{center}

\begin{abstract}
We study the LHC potential for discovering TeV-scale $SU(2)_L$ 5-plet
fermions introduced recently to explain small neutrino masses. We show that the Drell-Yan production and the decays of new exotic $\Sigma$ leptons are testable at the LHC. Their production is abundant due to nontrivial electroweak gauge charges. For $1\ \rm{fb^{-1}}$ of integrated luminosity at the present LHC $\sqrt{s}=7$ TeV, there can be 270 $\Sigma$-$\overline \Sigma$ pairs produced for $M_\Sigma=\rm{400\ GeV}$. Besides producing same-sign dilepton events, they could lead, due to a chosen small mixing between heavy and light leptons, to  $\sim$10 golden decays $\Sigma^{+++}(\overline{\Sigma^{+++}})\to W^{\pm} W^{\pm} l^{\pm}$ with a specific decay signature.
\end{abstract}

\vspace*{2 ex}

\begin{flushleft}
\end{flushleft}

\clearpage

\section{Introduction}

The start-up of the Large Hadron Collider (LHC) has enabled the exploration of previously inaccessible domains of heavy particles, which are at the core of various  extensions of the standard model (SM). A distinguished class of  modifications of the particle content of the SM attributes the smallness of the masses of observed neutrinos  to the presence of a seesaw mechanism. In such a setup
the neutrino masses can be realized through an effective dimension-five operator $LLHH$ \cite{Weinberg:1979sa}  which can be generated both at the tree and at the loop level. At the tree level there are three realizations of the dimension-five operator \cite{Ma:1998dn}, all of them based on an addition of a single new multiplet: Type-I \cite{Minkowski:1977sc-etc}, type-II \cite{KoK-etc} and type-III \cite{Foot:1988aq} seesaw mechanisms, mediated by heavy fermion singlet, scalar triplet and a fermion triplet, respectively. Thereby, the fermionic type-I and III models involve hypercharge zero heavy Majorana mediators.
Originally, the seesaw scale was linked to high energy GUT scale, while a recent reincarnation of $SU(5)$ GUT model
which within the adjoint 24 representation contains both a singlet and hypercharge zero fermion triplet, providing a low scale hybrid type-I and III seesaw model \cite{Bajc:2006ia}. In a multiple seesaw approach first put forward in \cite{Ma:2000cc} and later extended in \cite{Grimus:2009mm}, the involved additional fields and additional discrete symmetries
are instrumental in bringing the seesaw mechanism to the TeV scale \cite{Xing:2009hx} accessible at the LHC.

In the recently proposed approach of \cite{Picek:2009is}, which we are scrutinizing here, the lowering of the seesaw scale has been achieved without
use of discrete symmetries. By relying  only on the gauge symmetry and the renormalizability of the SM, we arrived at novel seesaw mechanism
supported by nonzero hypercharge fermions in higher isomultiplets.

Such a line of research follows from our examination \cite{Picek:2008dd}  of the vectorlike top partner, which by a sort of Dirac seesaw mechanism increases the mass of the top \cite{Vysotsky:2006fx}. We then extended our quark-sector study to the leptons, to explore a possible role of TeV-scale vectorlike Dirac fermions as seesaw mediators.

Indeed, the heaviness of the top quark and the lightness of the Higgs boson have been the landmarks for several extensions of the SM which rely on vectorlike rather than on sequential fourth generation fermions. We have observed that there are two options that such states, in
conjunction with appropriate new scalar fields, act as seesaw mediators: via the triplet fermions already introduced
in \cite{Babu:2009aq} and via the fermion 5-plet proposed in \cite{Picek:2009is}. The dimension nine (dim 9) operator produced at the tree level in the latter case  leads to the seesaw formula $m_{\nu} \sim v^6/ \Lambda_{NP}^5$, which reproduces the  empirical neutrino masses $m_{\nu} \sim 10^{-1}$ eV with sub-TeV 5-plet mass $M_\Sigma$. This calls to investigate  the phenomenology of new Dirac-type heavy leptons as the subject of our present study. Let us note that the focus of a study in Ref.~\cite{Babu:2009aq} with their exotic nonzero hypercharge triplet fermions has been on the accompanied triply-charged scalars, whereas a more recent study \cite{Liao:2010rx} considered the effects of these triplet fermions on the LFV decays of charged SM leptons. In addition, some further constraints on exotic triplets have been placed
very recently in Ref.~\cite{Chua:2010me,Delgado:2011iz}.

The present study of exotic  5-plet fermions can be compared to previous studies of hypercharge-zero triplets belonging to type-III seesaw mechanism: The earlier one performed by Franceschini et al. \cite{Franceschini:2008pz} and a subsequent by Li and He \cite{Li:2009mw}, which has some
overlap with more recent study \cite{Arhrib:2009mz} of type-III triplets appearing also in already mentioned hybrid type-I and III seesaw model \cite{Bajc:2006ia}.

This article is organized as follows. In Section 2 we describe a tree-level seesaw model based on nonzero hypercharge fermions and point out the nonminimal (conjunct) appearance of exotic fermions and scalars needed for dim 7 and dim 9 seesaw mechanisms. Section 3 deals with the production rates, and Section 4 with the subsequent decays of exotic 5-plet states. We summarize our results in Section 5. The details on the mixing between light and heavy leptons are exposed in the Appendix.

\section{ TeV-scale seesaw model with higher isomultiplets}
\label{petplet}

As a distinctive feature of the SM, the observed fundamental matter fields (quarks and leptons) are distributed in the lowest
possible representations of the SM group, $SU(3)_C \times SU(2)_L \times U(1)_Y$. In addition to the leptonic left-handed doublets and the right-handed singlets, there is also the SM Higgs doublet,
$H=(H^+, H^0)^T$. The values $(2T+1,Y)$ of the weak isospin $T$ and hypercharge $Y$, indicated in the parentheses,
\begin{eqnarray}\label{smfields}
&&L_L
\sim (2,-1), ~~~l_R\sim (1,-2),~~~H \sim (2,1),
\end{eqnarray}
reproduce the electric charges $Q=T_3+Y/2$ of the respective multiplet components.

Higher, isotriplet multiplets $\sim (3,2)$ and $\sim (3,0)$ are already familiar from type-II and III seesaw models, and it is obvious that we may encounter a 5-plet of zero hypercharge as the simplest generalization of type-III seesaw Majorana triplet to a higher-isospin multiplet.
Incidentally, such a $\sim (5,0)$ fermion multiplet has been singled out in the literature as a viable minimal dark matter (MDM)
candidate  \cite{Cirelli:2005uq}. It provides a completely neutral ($T_3=0$ and $Y=0$) component of an additional $SU(2)_L$ multiplet, and the minimality refers to its isolated appearance: MDM is realized as a single additional exotic matter particle coupled to the SM fields.\\
However, our concern here is with the seesaw models where exotic fermionic seesaw mediators appear together with appropriate exotic  scalar isomultiplets. Thereby, in order to add something new to already existing seesaw mechanisms, we employ
vectorlike fermionic multiplets with nonzero hypercharge. In  contrast to chiral SM fermions,
the masses of such vectorlike fermions are not restricted to the electroweak scale, so that they can be naturally
adjusted to the new physics scale. If in a novel seesaw mechanism the new physics appears already around the TeV scale,
then this scenario would not suffer from an additional, seesaw-induced hierarchy problem. In order to produce a tree-level seesaw, the newly introduced fermion multiplets have to have a neutral component which will mix with light neutrinos. Accompanying new scalar multiplets also
have to have a neutral component which can acquire the vacuum expectation value ({\em vev}).
As  exposed to more detail in the next subsection, these requirements together with the SM gauge symmetry lead to rather restricted viable quantum number assignments for the new multiplets: the one with vectorlike triplet fermions introduced in \cite{Babu:2009aq} leading to dim 7 seesaw operator, and the  vectorlike fermion 5-plet introduced by \cite{Picek:2009is} leading to dim 9 seesaw operator.

\subsection{Dim 9 seesaw model with fermionic 5-plets}

The effective operators of the form $(LLHH)(H^\dagger H )^n$ have already been studied in \cite{Gouvea:2008,Bonnet:2009ej}, and more recently in \cite{Liao:2010ku}, where it has been pointed out that the mass operator at each higher dimension is unique, independent of  the mechanism leading to it.

The model pursued here starts from three generations of SM model leptons represented by $L_L$ and $l_R$ in Eq.~(\ref{smfields}) and adds to them $n_\Sigma$ isospin $T=2$ vectorlike 5-plets of leptons with hypercharge two, where both left and right components transform as $(5,2)$ under the electroweak part of the SM group,
\begin{eqnarray}\label{fiveplet}
&&\Sigma_{L,R}=\left(\begin{array}{l}
\Sigma^{+++}\\\Sigma^{++}\\\Sigma^{+}\\\Sigma^{0}\\\Sigma^{-}
\end{array}\right)_{L,R}\sim~~(5,2) ~~.
\end{eqnarray}
Besides the SM Higgs doublet $H$ in Eq.~(\ref{smfields})
there are two additional isospin $T=3/2$ scalar multiplets $\Phi_1$
and $\Phi_2$
transforming as $(4,-3)$ and $(4,-1)$, respectively:
\begin{eqnarray}\label{fourplets}
&&\Phi_1=\left(\begin{array}{l}
\phi_1^0\\\phi_1^-\\\phi_1^{--}\\\phi_1^{---}
\end{array}\right)\sim~~(4,-3),~~~
\Phi_2=\left(\begin{array}{l}
\phi_2^+\\\phi_2^0\\\phi_2^{-}\\\phi_2 ^{--}
\end{array}\right)\sim~~(4,-1).~
\end{eqnarray}
Explaining the neutrino masses by dim 9 operator leads to the exotic matter shown in Eqs.~(\ref{fiveplet}) and (\ref{fourplets}). Concerning the fermion fields $\Sigma$, they should be vectorlike in order to avoid the chiral anomaly. The extra scalar fields are constrained by a requirement of the renormalizability of the extended SM, restricting the Lagrangian of the scalar sector to dimension 4 terms. Accordingly, the weak isospin of the exotic scalar fields ($\Phi_{1,2}$) can not be higher than three half, in order that they couple linearly with the Higgs field and develop an induced \textit{vev} after the spontaneous symmetry breaking of the SM gauge group. For this reason, the scalar fields should be in multiplets which contain electrically neutral components.

\begin{table}[h]
\begin{center}
\begin{tabular}{ccccc}
\hline
Seesaw Type    & Exotic Fermion           & Exotic Scalar                       & Scalar Coupling        & $m_\nu$ at     \\ \hline
Type-I         & $N_R\sim(1,0)$           & -                                      & -                     & dim 5          \\
Type-II        & -                        & $\Delta\sim(3,2)$                        & $\mu \Delta H H$        & dim 5          \\
Type-III       & $N_R\sim(3,0)$           & -                                      & -                     & dim 5          \\ \hline
Conjunct    & Exotic Fermion           & Exotic Scalars                       & Scalar - Higgs        & $m_\nu$ at     \\
    Mediator       &  Pair                    & $\Phi_1,\ \Phi_2$                   & Couplings        &      \\ \hline
doublet        & $\Sigma_{L,R}\ (2,1)$  & $(3,-2),\   (3,0)$  & $\mu_{1,2} \Phi_{1,2} H H$ & dim 5          \\
triplet        & $\Sigma_{L,R}\ (3,2)$  & $(4,-3),\   (2,-1)$   & $\lambda_1\Phi_1 H H H$       & dim 7          \\
4-plet      & $\Sigma_{L,R}\ (4,1)$  & $(3,-2),\   (3,0)$  & $\mu_{1,2} \Phi_{1,2} H H$ & dim 5          \\
5-plet      & $\Sigma_{L,R}\ (5,2)$  & $(4,-3),\   (4,-1)$   & $\lambda_{1,2}\Phi_{1,2} H H H$   & dim 9          \\\hline
\end{tabular}
\caption{The assignments of electroweak charges for exotic particles leading to the tree-level seesaw operator up to dim 9.}
\label{tableI}
\end{center}
\end{table}

Since the sought-after dim 9 operator can be relevant only in the absence of possible dim 5 operators, we forbid the appearance of those states which generate conventional seesaw mechanisms. In first place, we should avoid a scalar triplet $\Delta\sim(3,2)$
generating type-II seesaw, and a fermion singlet $N_R\sim(1,0)$ or triplet $N_R\sim(3,0)$ generating
type-I and III seesaw, respectively (see the Table 1). Note that type-I, II, and III mechanisms correspond to already mentioned minimal (single) appearance of exotic (EX) scalar/fermionic particles in the Yukawa term
\begin{equation}\label{minYuk}
\mathcal{L}_{Y}^\text{minimal}\sim \text{(SM-fermion) (SM-scalar/fermion) (EX-fermion/scalar)}.
\end{equation}
Also, a recent study of exotic triplets  in Ref.~\cite{Chua:2010me,Delgado:2011iz} restricts to such minimal Yukawa term. We generalize Eq.~(\ref{minYuk}) to new Yukawa terms with double appearance of exotic particles,
\begin{equation}
\mathcal{L}_{Y}^\text{conjunct} \sim \text{(SM-fermion) (EX-scalar) (EX-fermion)}\ ,
\label{conj}
\end{equation}
where a conjunct exotic scalar-fermion pair can  lead to the neutrino mass operators of dimension larger then 5, as shown in Table \ref{tableI}.

\begin{figure}[h]
\begin{center}
\includegraphics[scale=1.2]{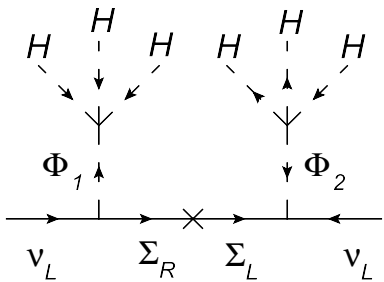} \hspace{0.7cm}
\end{center}
\caption{Dim 9 tree-level seesaw diagram produced by exotic fields in Eqs.~(\ref{fiveplet}) and (\ref{fourplets}). }
\label{tree-diagrams}
\end{figure}

Consequently, the focus here is on nonzero hypercharge Dirac fermions $\Sigma_{L,R}$ from the last row of Table \ref{tableI},
playing role of the seesaw mediators which produce the mechanism displayed on Fig.~\ref{tree-diagrams}.
The seesaw diagram on this figure arises due to the Yukawa couplings explicated in
Eq.~(\ref{gaugelag}) and the scalar-field couplings $\lambda_1$ and $\lambda_2$  explicated in Eq.~(\ref{pot}). Note that it involves a decomposition of the appropriate couplings in terms of the Clebsch-Gordan coefficients.

In order to avoid already known seesaw mechanisms, there is a gap between the quantum numbers of the SM fields  and the EX fields. To overbridge this gap and to obtain the gauge invariant interaction in Eq.~(\ref{conj}), the conjunct EX scalar-fermion pair should form an object with the weak isospin half and the hypercharge one. One possible way to achieve this is shown in the doublet and 4-plet ``Seesaw Type`` rows in Table~\ref{tableI}, where new fermion fields of isospin one half (doublet) or three half (4-plet) hold together with a scalar of isospin one (triplet). However the latter appears already in type-II seesaw and would lead to dim 5 operator for neutrino masses. Then we are left with new fermion fields which should be of either isospin one (triplet) or isospin two (5-plet). Whereas the first option corresponds to dim 7 operator introduced by Babu et al. \cite{Babu:2009aq}, we adopt the latter case corresponding to dim 9 operator which might be more accessible for tests at the LHC. The requirement of employing the Dirac fermions as tree-level seesaw mediators restricts their hypercharge to $Y=2$ for both the triplet and the 5-plet case.

\subsection{Mass matrices of the 5-plet states}

For our choice of EX fields in Eqs.~(\ref{fiveplet}) and (\ref{fourplets}) the conjunct Yukawa interaction in Eq.~(\ref{conj}) is explicated by the second row in the following gauge invariant Lagrangian:
\begin{eqnarray}\label{gaugelag}
\mathcal{L}&=& \overline{\Sigma_L}i\cancel{D}\Sigma_L + \overline{\Sigma_R}i\cancel{D}\Sigma_R
-\overline{\Sigma_R}M_\Sigma \Sigma_L -\overline{\Sigma_L}M_\Sigma^\dagger \Sigma_R \nonumber \\
&+& \left( \overline{\Sigma_R}Y_1L_L\Phi_1^* + \overline{(\Sigma_L)^c} Y_2 L_L \Phi_2+ \mathrm{H.c.} \right)\ .
\end{eqnarray}
This contains the bare Dirac mass terms for the introduced  vectorlike lepton 5-plet fields
\begin{eqnarray}
-\mathcal{L}_D = && \overline{\Sigma_R^{+++}} M_\Sigma \Sigma_L^{+++} + \overline{\Sigma_R^{++}} M_\Sigma \Sigma_L^{++} \nonumber \\
&& + \overline{\Sigma_R^+} M_\Sigma \Sigma_L^+ + \overline{\Sigma_R^0} M_\Sigma \Sigma_L^0 + \overline{\Sigma_R^-} M_\Sigma \Sigma_L^- + \mathrm{H.c.}  \ .\label{Dirmass}
\end{eqnarray}
Additional mass terms arise from the Yukawa part in Eq.~(\ref{gaugelag}). As discussed shortly, an analysis of the scalar potential
shows that the neutral components of two scalar quadruplets $\Phi^0_1$ and $\Phi^0_2$ develop the induced {\em vevs}, $v_{\Phi_1}$ and $v_{\Phi_2}$, respectively. This means that Yukawa terms in Eq.~(\ref{gaugelag}) lead to the mass terms connecting the SM lepton doublet with new vectorlike lepton 5-plet. Thereby, the triply and doubly charged states do not mix with SM leptons whereas the singly-charged and neutral states have peculiar mixing with SM leptons which we consider in detail in the Appendix.

When extracting the mass matrices produced by the mass and Yukawa terms we suppress the generation indices of the fields. Three neutral left-handed fields $\nu_L$, $\Sigma_L^{0}$ and $(\Sigma_R^{0})^c$ span the
symmetric neutral mass matrix as follows:
\begin{eqnarray}
\mathcal{L}_{\nu \Sigma^0} =  \, -\frac{1}{2}
\left(  \overline{(\nu_L)^c}  \; \overline{(\Sigma_L^0)^c} \; \overline{\Sigma_R^0} \right)
\left( \! \begin{array}{ccc}
0 & m_2^T & m_1^T \\
m_2 & 0 & M_{\Sigma}^T \\
m_1 & M_{\Sigma} & 0
\end{array} \! \right) \,
\left( \!\! \begin{array}{c} \nu_L \\ \Sigma_L^0 \\ (\Sigma_R^0)^c \end{array} \!\! \right)
\; + \mathrm{H.c.}\ .
\label{neutral_mass_matrix}
\end{eqnarray}
For singly-charged fermions we arrive at the nonsymmetric mass matrix which explicates the mixing between the SM leptons and new singly-charged states
\begin{eqnarray}
\mathcal{L}_{l\Sigma} =  \,
- \left( \overline{l_R} \; \overline{\Sigma_R^-} \; \overline{(\Sigma_L^+)^c} \right)
\left( \! \begin{array}{ccc}
m_l & 0 & 0 \\
m_3 & M_{\Sigma} & 0 \\
m_4 & 0 & M_{\Sigma}^T
\end{array} \! \right) \,
\left( \!\! \begin{array}{c} l_L \\ \Sigma_L^- \\ (\Sigma_R^+)^c \end{array} \!\! \right)
\; + \mathrm{H.c.} \ .
\label{charged_mass_matrix}
\end{eqnarray}
Here $M_{\Sigma}$ is given in Eq.~(\ref{Dirmass}), while $m_3$ and $m_4$ entries in Eq.~(\ref{charged_mass_matrix}) and $m_1$ and $m_2$ in the neutral sector in Eq.~(\ref{neutral_mass_matrix}) come from the respective terms in Eq.~(\ref{gaugelag}):
\begin{eqnarray}\label{masses}
m_1 = \sqrt{\frac{1}{10}}Y_1{v^*_\Phi}_1\ , &&\qquad m_2 = - \sqrt{\frac{3}{20}} Y_2{v_\Phi}_2\ , \nonumber \\
m_3 = \sqrt{\frac{2}{5}}Y_1{v^*_\Phi}_1\ , &&\qquad m_4 = \sqrt{\frac{1}{10}} Y_2{v_\Phi}_2\ .
\end{eqnarray}
Accordingly, the masses $m_i$ ($i = 1, 2, 3, 4$) are determined by the {\em vev} $v_{\Phi_1}$ and {\em vev} $v_{\Phi_2}$ of the neutral components of the scalar quadruplets, whereas $M_{\Sigma}$ is on the new physics scale $\Lambda_{NP}$, which is larger than the electroweak scale.

The diagonalization of the mass matrices proceeds by unitary transformations presented in the Appendix. They are the result of a procedure
developed in Ref.~\cite{Grimus:2000vj}, whereby we write the leading order expressions up to $M^{-2}_\Sigma$ in the basis where the matrices $m_l$ and $M_\Sigma$ are already diagonalized. Block diagonalization of the neutral mass matrix in Eq.~(\ref{neutral_mass_matrix}) gives to this order
\begin{equation}\label{effseesaw}
    \tilde{m}_\nu\simeq - m_2^T M^{-1}_\Sigma m_1 - m_1^T M^{-1}_\Sigma m_2 \ ,
\end{equation}
where a familiar unitary $V(3 \times 3)_{PMNS}$ matrix diagonalizes the obtained effective light neutrino mass matrix:
\begin{equation}\label{effPMNS}
 V_{PMNS}^T \tilde{m}_\nu V_{PMNS} = m_\nu \ .
\end{equation}
The same formalism has been applied in Refs.~\cite{Li:2009mw,Hettmansperger:2011bt}
and the results in our Appendix agree with theirs, when taking the appropriate limits.

\subsection{Dim 9 w.r.t.  dim 5 loop-level seesaw mechanisms}

Turning to the scalar potential, we restrict ourselves to already mentioned renormalizable terms, relevant for our mechanism:
\begin{eqnarray}\label{pot}
V(H, \Phi_1, \Phi_2) &\sim& \mu_H^2 H^\dagger H + \mu^2_{\Phi_1} \Phi^\dagger_1 \Phi_1+ \mu^2_{\Phi_2} \Phi^\dagger_2 \Phi_2 + \lambda_H (H^\dagger H )^2 \nonumber \\
 &+& \{ \lambda_1 \Phi^*_1 H^* H^* H^* + \mathrm{H.c.} \} + \{ \lambda_2 \Phi^*_2 H H^* H^* + \mathrm{H.c.} \} \nonumber\\
 &+& \{ \lambda_3 \Phi^*_1 \Phi_2 H^* H^* + \mathrm{H.c.} \} \ .
\end{eqnarray}
The electroweak symmetry breaking proceeds in the usual way from the {\em vev} $v=174\ \rm{GeV}$ of the Higgs doublet, implying $\mu_H^2<0$. On the other hand, the electroweak $\rho$ parameter dictates the {\em vevs} $v_{\Phi_1}$ and $v_{\Phi_2}$ to be small, implying $\mu^2_{\Phi_1},\mu^2_{\Phi_2}>0$. However, the $\lambda_1$ and $\lambda_2$ terms in  Eq.~(\ref{pot}) result in the induced {\em vevs} for the isospin 3/2 scalar multiplets,
\begin{equation}\label{phivev}
    v_{\Phi_1} \simeq -\lambda_1 \frac{v^3}{\mu^2_{\Phi_1}} \, \, ,\, \, 
v_{\Phi_2} \simeq -\lambda_2 \frac{v^3}{\mu^2_{\Phi_2}} \, \, .
\end{equation}
Let us stress that we allow only the SM scalar doublet  to wear a Mexican hat, and that the terms linear
in the extra scalar fields, which induce their {\em vevs}, only slightly deform the total scalar potential without affecting
the {\em vev} of the SM Higgs. Generically, the potential in Eq.~(\ref{pot}) is at large field values dominated
by the terms quartic in fields, and with a number of free parameters in the potential there is no problem to ensure the stability of the SM vacuum.

The nonanishing {\em vevs} in Eq.~(\ref{phivev}) change the electroweak $\rho$ parameter to $\rho (\Phi_1) \simeq 1-6v^2_{\Phi_1}/v^2$ and $\rho (\Phi_2) \simeq 1+6v^2_{\Phi_2}/v^2$, respectively. If these {\em vevs} are taken separately, the experimental value $\rho=1.0000^{+0.0011}_{-0.0007}$ \cite{Nakamura:2010zzi} leads to the upper bounds $v_{\Phi_1}\leq1.9$ GeV and $v_{\Phi_2}\leq2.4$ GeV, respectively. Thus, the value of a few GeV can be considered as an upper bound on both $v_{\Phi_1}$ and $v_{\Phi_2}$ if there is no fine-tuning between them. By merging
Eqs.~(\ref{masses}), ~(\ref{effseesaw}) and ~(\ref{phivev}) we obtain for the light neutrino mass
\begin{equation}\label{dim9}
    m_{\nu} \sim \frac {Y_1 Y_2\ \lambda_1\lambda_2\ v^6} {M_{\Sigma}\ \mu^2_{\Phi_1}\ \mu^2_{\Phi_2}} \,,
\end{equation}
which leads to dim 9 seesaw formula result, $m_\nu \sim v^6/\Lambda_{NP}^5$.
\\

\begin{figure}[t]
\begin{center}
\includegraphics[scale=1.1]{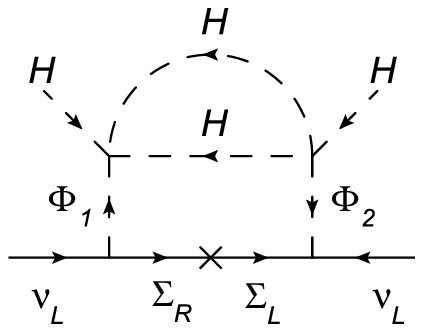} \hspace{0.7cm}
\includegraphics[scale=1.1]{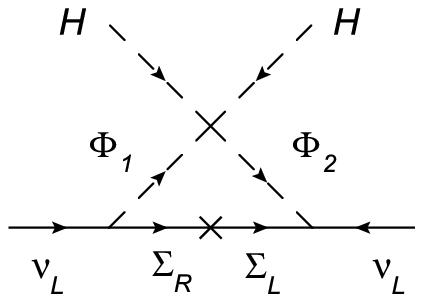}
\end{center}
\caption{Dim 5 operator arising at the loop level: the two-loop seesaw diagram given by
the $\lambda_1$ and $\lambda_2$ couplings in  Eq.~(\ref{pot}), and
the one-loop seesaw diagram given by the $\lambda_3$ coupling in  Eq.~(\ref{pot}).}
\label{loop-diagrams}
\end{figure}

Neutrino masses can arise from dim 5 operators generated at the loop level. The loop factors provide a sufficient suppression as exploited by one- \cite{Zee:1980ai}, two- \cite{Zee:1985id,Babu:1988ki} and three-loop level \cite{Krauss:2002px} mechanisms. Such a loop level generated dim 5 operator arises also in our extension of the SM. As explained in \cite{Picek:2009is}, there is a restricted range of the parameter space where the dim 5 loop-contributions shown on Fig.~\ref{loop-diagrams} are smaller than the dim 9 tree-level contribution shown on Fig.~\ref{tree-diagrams}. With an extra assumption on the scalar field coupling strengths in  (\ref{pot}), $\lambda_3 \simeq \lambda_1 \cdot \lambda_2$, the dim 9 mechanism turns out to be the leading one for $\Lambda_{NP} \sim$ few 100 GeV. The loop
contributions prevail for $\Lambda_{NP} \sim$ several TeV, which are beyond the direct reach of the LHC, and may also mark a virtual physics in the LHC era.

\section{Production of 5-plet leptons at the LHC}
\label{sec:production}

The production channels of the  heavy 5-plet leptons in proton-(anti)proton collisions are
dominated by the quark-antiquark annihilation via neutral and charged gauge bosons
\begin{displaymath}
q + \bar{q} \to A \to \Sigma + \bar{\Sigma}\;, \qquad A = \gamma, Z, W^\pm \;,
\end{displaymath}
where the gauge Lagrangian relevant for the production
\begin{eqnarray}\label{gauge-short}
\mathcal{L}_{gauge}^{\Sigma \overline{\Sigma}}=
&+& e (3\overline{\Sigma^{+++}}\gamma^\mu \Sigma^{+++} + 2\overline{\Sigma^{++}}\gamma^\mu \Sigma^{++}
+ \overline{\Sigma^{+}}\gamma^\mu \Sigma^{+} - \overline{\Sigma^{-}}\gamma^\mu \Sigma^{-})A_\mu \nonumber \\
&+&{g \over c_W}( (2-3s_W^2) \overline{\Sigma^{+++}}\gamma^\mu \Sigma^{+++} + (1-2s_W^2) \overline{\Sigma^{++}}\gamma^\mu\Sigma^{++})Z_\mu\\
&+&{g \over c_W}( (-s_W^2) \overline{\Sigma^{+}}\gamma^\mu \Sigma^{+} + (-1) \overline{\Sigma^{0}}\gamma^\mu\Sigma^{0} + (-2+s_W^2) \overline{\Sigma^{-}}\gamma^\mu\Sigma^{-})Z_\mu\nonumber\\
&+&g(\sqrt{2}\overline{\Sigma^{++}}\gamma^\mu\Sigma^{+++} + \sqrt{3}\overline{\Sigma^{+}}\gamma^\mu\Sigma^{++}
+ \sqrt{3}\overline{\Sigma^{0}}\gamma^\mu\Sigma^{+} + \sqrt{2}\overline{\Sigma^{-}}\gamma^\mu\Sigma^{0}) W^-_\mu + \mathrm{H.c.}\nonumber
\end{eqnarray}
is contained in Eq.~(\ref{gauge-complete}) in the Appendix.
The cross section for the partonic process is determined entirely by gauge couplings and is given by
\begin{equation}
\hat{\sigma} (q \bar{q} \to \Sigma \overline{\Sigma}) =
\frac{\beta (3-\beta^2)}{48 \pi} \hat{s} (V_{L}^2 + V_{R}^2) \;,
\label{eq:xspartonic}
\end{equation}
where $\hat{s}\equiv(p_q + p_{\bar{q}})^{2}$ is the Mandelstam variable $s$ for the
quark-antiquark system, the parameter $\beta\equiv\sqrt{1-4 M_\Sigma^2/\hat{s}}$ denotes the
heavy lepton velocity, and the left- and right-handed couplings in Eq.~(\ref{eq:xspartonic}) are given by
\begin{align}
V_{L,R}^{(\gamma+Z)}&= \frac{Q_{\Sigma} Q_q e^2}{\hat{s}} +
    \frac{g^{Z\Sigma}\, g_{L,R}^{q} \, g^2}{c_{W}^2(\hat{s}-M_{Z}^2)}\;, \\
V_{L}^{(W^-)}&= \frac{g^{W\Sigma} g^2 V_{ud}}{\sqrt{2}(\hat{s}-M_{W}^2)} =
  V_{L}^{(W^+)^*}\;,  \\
V_{R}^{(W^\pm)}&= 0 \;.
\end{align}
Here,
$g^{q}_L = T_3 - s_{W}^2 Q_q $ and $g^{q}_R = - s_{W}^2 Q_q$
are the SM chiral quark couplings to the $Z$ boson.
The vector couplings of heavy leptons to gauge bosons are
\begin{equation}\label{vectorcouplings}
g^{Z\Sigma} = T_3 - s_{W}^2 Q_\Sigma\ ,\ g^{W\Sigma} = \sqrt{2}, \sqrt{3}, \sqrt{3}\ \rm{and}\ \sqrt{2}\ ,
\end{equation}
where $g^{W\Sigma}$ can be read of the last raw in Eq.~(\ref{gauge-short})
relevant for the production of
$\Sigma^{+++}\overline{\Sigma^{++}}$,
$\Sigma^{++}\overline{\Sigma^{+}}$,
$\Sigma^{+}\overline{\Sigma^{0}}$ and
$\Sigma^{0}\overline{\Sigma^{-}}$ pairs, respectively.
We note, in passing, that the choices $g^{Z\Sigma}=c_{W}^2$ and
$g^{W\Sigma}=1$, appropriate for the type-III seesaw model, reproduce the expressions given in \cite{Franceschini:2008pz}.

\begin{figure}[h]
\begin{center}
\includegraphics[scale=0.70]{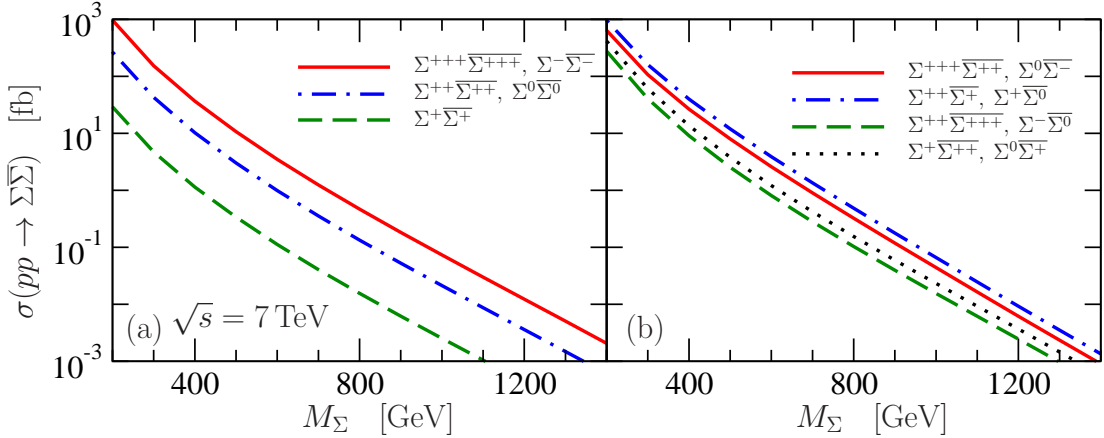}
\end{center}
\caption{The cross sections for production of 5-plet lepton  pairs on LHC
proton-proton collisions at $\sqrt{s}=7\, {\rm TeV}$ via
neutral $\gamma, Z$ (a) and charged $W^{\pm}$ currents (b),
in dependence on the heavy 5-plet mass $M$.}
\label{fig:LHCproduction}
\end{figure}

To obtain the cross sections for a hadron collider, the partonic cross-section
(\ref{eq:xspartonic}) has to be convoluted with the appropriate parton distribution
functions (PDF) $q(x, \mu^2)$. For the case of the proton-proton collisions we
have
\begin{multline}
\frac{d^2 \sigma(pp \to \Sigma \overline{\Sigma})}{d x_1 d x_2} =
\frac{1}{N_c} \sum_{f_{1}=u,d,\dots} \sum_{f_{2}=\bar{u},\bar{d},\dots}
\:
\hat{\sigma} (q_{f_{1}} {q}_{f_{2}} \to \Sigma \overline{\Sigma})
\\ \times
\big( q_{f_{1}}(x_1, \mu^2) q_{f_{2}}(x_2, \mu^2) + q_{f_{2}}(x_1,
\mu^2)
q_{f_{1}}(x_2, \mu^2) \big) \;,
\end{multline}
leading to the total cross section
\begin{multline}
\sigma(pp \to \Sigma \overline{\Sigma}) =
\frac{1}{N_c} \sum_{f_{1}=u,d,\dots}
\sum_{f_{2}=\bar{u},\bar{d},\dots}
\int_{4 M^2/s}^1 d \tau\, \int_{\tau}^1 \frac{d x_1}{x_1} \:
\hat{\sigma} (q_{f_{1}} {q}_{f_{2}} \to \Sigma \overline{\Sigma})
\\ \times
\Big( q_{f_{1}}(x_1, \mu^2) q_{f_{2}}(\frac{\tau}{x_1}, \mu^2) +
q_{f_{2}}(x_1, \mu^2)
q_{f_{1}}(\frac{\tau}{x_1}, \mu^2) \Big) \;.
\label{eq:xshadronic}
\end{multline}
Here, the inverse of the $N_c=3$ factor is due to the color averaging in the initial state, corresponding to the
summation over colors implicit in the standard published PDF functions.
To evaluate the cross sections we have used CTEQ6.6 PDFs \cite{Nadolsky:2008zw}
via LHAPDF software library \cite{Whalley:2005nh} and
have chosen $\mu=M_\Sigma$ as a factorization scale.
Comparison with other PDF sets has shown that uncertainty in
cross sections induced by the choice of PDFs is less than 3\%.
We find an excellent agreement when checking our calculations with an independent evaluation of
cross sections performed by using MadGraph \cite{Alwall:2007st}, where the
present model has been implemented with the help of FeynRules
package \cite{Christensen:2008py}.

\begin{figure}[h]
\begin{center}
\includegraphics[scale=0.70]{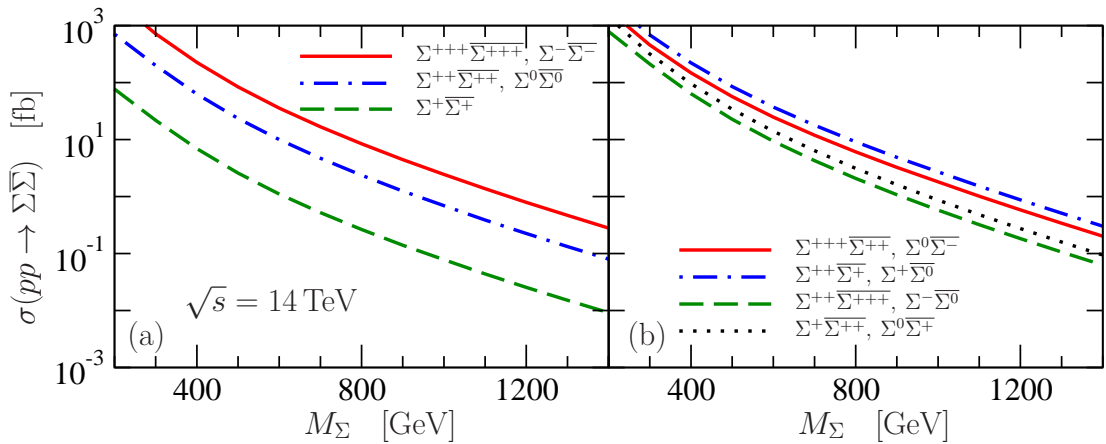}
\end{center}
\caption{Same as Fig.~\ref{fig:LHCproduction}, but for designed  $\sqrt{s}=14\, {\rm TeV}$ at the LHC.}
\label{fig:production14}
\end{figure}

\begin{figure}[h]
\begin{center}
\includegraphics[scale=0.70]{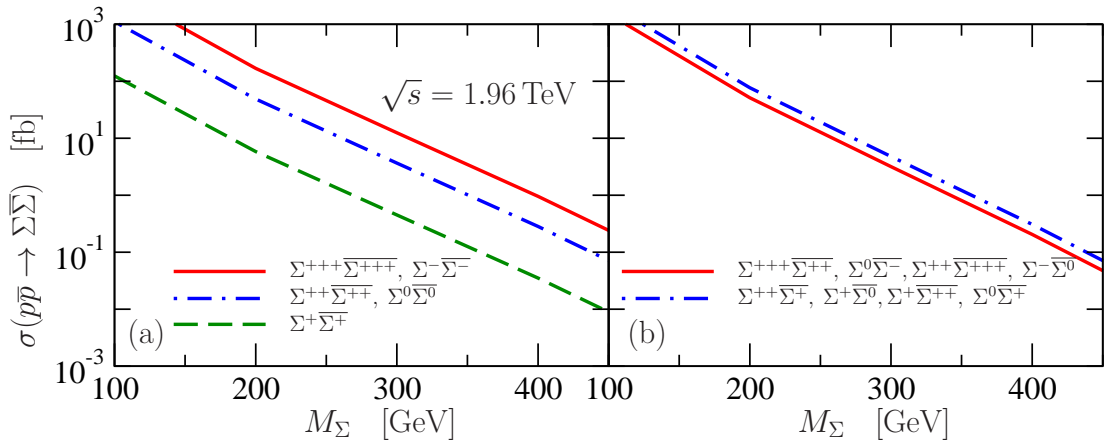}
\end{center}
\caption{Same as Fig.~\ref{fig:LHCproduction}, but for $\sqrt{s}=1.96\,{\rm TeV}$
at the Tevatron.}
\label{fig:productionTEV}
\end{figure}

The cross sections for proton-proton collisions are presented at $\sqrt{s} = 7\, {\rm TeV}$,
appropriate for the current LHC run, on
Fig.~\ref{fig:LHCproduction}, and for designed $\sqrt{s}=14\,{\rm TeV}$ on Fig.~\ref{fig:production14}. On Fig.~\ref{fig:productionTEV} cross sections for proton-antiproton collisions, relevant for the Tevatron at $\sqrt{s}=1.96\,{\rm TeV}$ are presented. Thereby we distinguish separately the production via neutral currents shown on LHS, and via charged currents shown on RHS of Figs.~\ref{fig:LHCproduction} to \ref{fig:productionTEV}.

\begin{table}[ht]
\begin{center}
\begin{tabular}{cccc}
\hline
Produced                     &                & Cross section ($\rm{fb}$)  &              \\
pair                        & $M_\Sigma = 200\ \rm{GeV}$  & $M_\Sigma = 400\ \rm{GeV}$  &  $M_\Sigma = 800\ \rm{GeV}$ \\ \hline
$\Sigma^{+++} \overline{\Sigma^{+++}}$   & 967            & 36.6               & 0.47                             \\
$\Sigma^{++} \overline{\Sigma^{++}}$     & 267            & 10.3               & 0.13                             \\
$\Sigma^{+} \overline{\Sigma^{+}}$       & 29            & 1.1                & 0.02                             \\
$\Sigma^{0} \overline{\Sigma^{0}}$       & 253            & 9.3                & 0.11                             \\
$\Sigma^{-} \overline{\Sigma^{-}}$       & 939            & 34.6               & 0.43                             \\ \hline
$\Sigma^{+++} \overline{\Sigma^{++}}$    & 641            & 26.5               & 0.32                             \\
$\Sigma^{++} \overline{\Sigma^{+}}$      & 961           & 39.8                & 0.49                             \\
$\Sigma^{+} \overline{\Sigma^{0}}$       & 961            & 39.8                & 0.49                             \\
$\Sigma^{0} \overline{\Sigma^{-}}$       & 641            & 26.5               & 0.32                             \\ \hline
$\Sigma^{++} \overline{\Sigma^{+++}}$    & 276            & 9.1                & 0.10                              \\
$\Sigma^{+} \overline{\Sigma^{++}}$      & 414           & 13.6               & 0.16                             \\
$\Sigma^{0} \overline{\Sigma^{+}}$       & 414            & 13.6               & 0.16                             \\
$\Sigma^{-} \overline{\Sigma^{0}}$       & 276            & 9.1                & 0.10                             \\ \hline
Total                                    & 7040           & 269.9             & 3.30                              \\ \hline
\end{tabular}
\caption{Production cross sections for $\Sigma$-$\overline{\Sigma}$ pairs for the current LHC run at $\sqrt{s} = 7\, {\rm TeV}$, for three selected values of $M_\Sigma$.}
\label{tableII}
\end{center}
\end{table}

We extract the pair production cross sections shown in Fig.~\ref{fig:LHCproduction} for three selected values of $M_\Sigma$ which
we show in Table \ref{tableII}.
From this table we find that the triply-charged $\Sigma^{+++}$ has the biggest production cross section $\sigma(\Sigma^{+++})|_{M_\Sigma = 400 \rm{GeV}}= 63.1\, \rm{fb}$ and singly-charged $\Sigma^{-}$ has the smallest, but comparable, production cross section $\sigma(\Sigma^{-})|_{M_\Sigma = 400 \rm{GeV}}= 43.7\, \rm{fb}$. Production cross sections for all other particles and antiparticles from a 5-plet are in between. In particular, for $1\ \rm{fb^{-1}}$ of integrated luminosity at the present LHC $\sqrt{s}=7$ TeV, there would be 270 $\Sigma$-$\overline \Sigma$ pairs in total produced for $M_\Sigma=400 \,\rm{GeV}$, among which are $\sim 100$ triply-charged $\Sigma^{+++}$ or $\overline{\Sigma^{+++}}$ fermions. Note that the production at the LHC and  Tevatron, which is comparable in the case of light $\Sigma$-states,  changes dramatically in favor of the LHC for higher values of the $\Sigma$-lepton masses. Therefore, only the restricted  mass region is covered on Fig.~\ref{fig:productionTEV} for the Tevatron.

By testing the heavy lepton production cross sections one can hope to identify the quantum numbers of the 5-plet particles,
but in order to confirm their relation to neutrinos one has to study their decays.

\section{Decays of 5-plet states}
\label{sec:triplychargeddecay}

Provided that in our scenario the exotic scalar states in Eq.~(\ref{fourplets}) are slightly heavier than the exotic leptons in Eq.~(\ref{fiveplet}), the exotic scalars will not appear in the final states in  heavy lepton decays. Also, the most phenomenologically interesting triply-charged scalar state is common to our model and the model by Babu et al.~\cite{Babu:2009aq} where it has been in the focus of their study, so that there is no need to repeat their study  here. Thus we focus entirely on the decay modes of the heavy lepton states in Eq.~(\ref{fiveplet}).
Let us note that the ordering of the states in this formula corresponds to the spectrum of these states, running from the lightest $\Sigma^-$
to the heaviest $\Sigma^{+++}$ state.

The mass difference induced by loops of SM gauge bosons between two components of $\Sigma$ 5-plet with electric charges $Q$ and $Q'$ is explicitly calculated in \cite{Cirelli:2005uq}
\begin{eqnarray}\label{massdifferences}
M_Q -  M_{Q'} =&&\frac{\alpha_2 M}{4\pi} \Big\{ (Q^2-Q^{\prime 2})s_{\rm W}^2 f(\frac{M_Z}{M})+(Q-Q')(Q+Q'-Y) \nonumber \\  && \bigg[f(\frac{M_W}{M})-f(\frac{M_Z}{M})\bigg] \Big\}\ ,
\end{eqnarray}
where
\begin{equation}
f(r) = {r \over 2}  \left[2 r^3\ln r -2 r+(r^2-4)^{1/2} (r^2+2) \ln \left( {r^2 -2 - r\sqrt{r^2-4} \over 2} \right)\right] \ .
\end{equation}
For $M_{\Sigma}=400\ \rm{GeV}$ we get
\begin{eqnarray}\label{splitting}
&&M_{\Sigma^{+++}} - M_{\Sigma^{++}} \simeq 1130\ \rm{MeV},\ M_{\Sigma^{++}} - M_{\Sigma^{+}} \simeq 804\ \rm{MeV},\nonumber\\
&&M_{\Sigma^{+}} - M_{\Sigma^{0}} \simeq 477\ \rm{MeV},\ M_{\Sigma^{0}} - M_{\Sigma^{-}} \simeq 150\ \rm{MeV}\ .
\end{eqnarray}
The neutral $\Sigma^0$ and singly-charged $\Sigma^+$ states, coming in Eq.~(\ref{splitting}), receive additional corrections given in Eqs.~(\ref{neutralcorr}) and (\ref{Mrazvoj}), respectively. However, being of the order of magnitude of the light neutrino masses, these corrections are
completely negligible.

The splitting $\sim 1$ GeV between triply and doubly charged 5-plet states opens additional decay channels, in particular, the $\rho$ resonance enhanced decay channel to the two pions in the final state\index{\footnote{}}.

\subsection{Pointlike $M_\Sigma$-dependent decays}

The Lagrangian in the mass-eigenstate basis relevant for the decays of the heavy leptons is
the result of a detailed calculation exposed in the Appendix. Let us repeat here the neutral
and charged current terms,
\begin{eqnarray}
  \mathcal{L}_{NCZ} &=& {g\over
c_W}\Big[ \overline{\nu} ( {3\over2} V_{PMNS}^\dagger V_1 \gamma^\mu P_L
-{\sqrt 3 \over 2 \sqrt 2} V_{PMNS}^T V^*_2 \gamma^\mu P_R ) \Sigma^0 \nonumber \\
&+& \overline{l} (3V_1 \gamma^\mu P_L )\Sigma^- + \overline{l^c} ({1 \over 2} V^*_2  \gamma^\mu P_R )
\Sigma^+ \Big] Z_\mu^0 + \mathrm{H.c.}\ ,
\end{eqnarray}
and
\begin{eqnarray}
\mathcal{L}_{CC} &=& g\Big[\overline{\nu}(- \sqrt 3 V_{PMNS}^\dagger V_1 \gamma^\mu P_L + \sqrt 2 V_{PMNS}^T V^*_2 \gamma^\mu P_R)\Sigma^+\nonumber\\
&+& \overline{\Sigma^-} (\sqrt 3 V^T_2 V_{PMNS}^* \gamma^\mu P_R) \nu + \overline{l} (-{3\over \sqrt 2} V_1 \gamma^\mu P_L) \Sigma^0 \nonumber\\
&+& \overline{\Sigma^0} (-{\sqrt 3\over 2} V^T_2 \gamma^\mu P_R) l^c + \overline{l^c} (-\sqrt 3 V^*_2 \gamma^\mu P_R) \Sigma^{++} \Big] W_\mu^- + \mathrm{H.c.}\
\end{eqnarray}
Here, all the couplings can be expressed  in terms of the matrix-valued quantities  $V_1$ and $V_2$ explicated in
the Appendix, corresponding to the mass matrices
in Eq.~(\ref{masses})
\begin{equation}
    V_1= \sqrt{\frac{1}{10}} Y^\dagger_1{v_\Phi}_1 M^{-1}_\Sigma\ , V_2= \sqrt{\frac{1}{10}} Y^\dagger_2{v^*_\Phi}_2 M^{-1}_\Sigma \ .
\end{equation}

Out of the five $\Sigma$-states in Eq.~(\ref{fiveplet}),
the four lying lowest have the pointlike decays to the gauge bosons  and  SM leptons.
Here, in contrast to  type-III triplet states, the decays of 5-plet states into the SM Higgs boson are suppressed. When appropriate,
we will comment on further differences with respect to the decays of type-III triplet states.\\

\begin{figure}
\begin{center}
\includegraphics[scale=0.70]{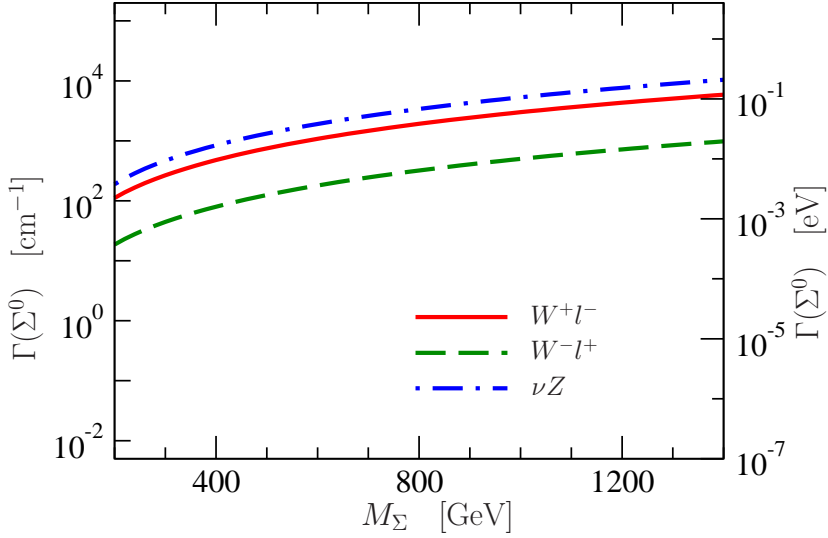}
\end{center}
\caption{Partial decay widths of $\Sigma^0$
5-plet lepton for $|V^{l\Sigma}_1|=|V^{l\Sigma}_2| = 10^{-6} \sqrt{{20 \over M_{\Sigma} \rm{(GeV)}}}$, in dependence
on heavy 5-plet mass $M_{\Sigma}$.}
\label{fig:decay0800}
\end{figure}

\begin{figure}
\begin{center}
\includegraphics[scale=0.70]{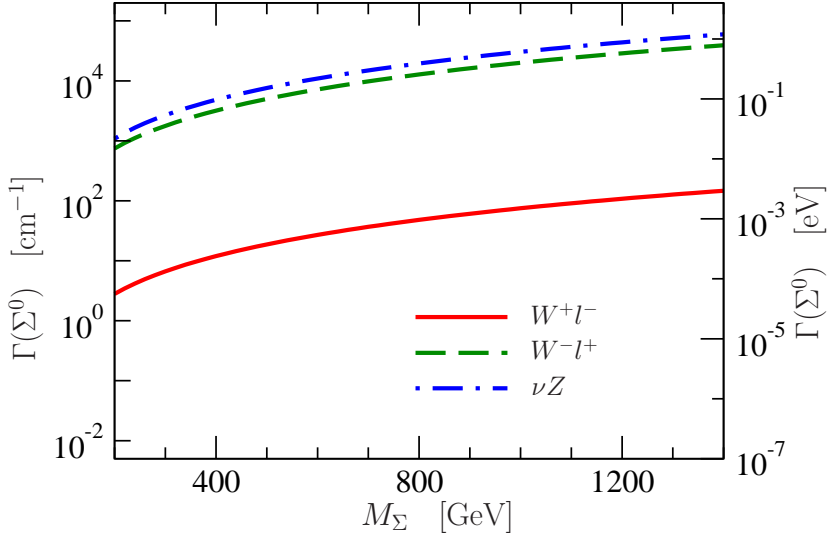}
\end{center}
\caption{Partial decay widths of $\Sigma^0$
5-plet lepton for $|V^{l\Sigma}_1| = 10^{-6} \sqrt{{0.5 \over M_{\Sigma} \rm{(GeV)}}}$ different from $|V^{l\Sigma}_2| = 10^{-6} \sqrt{{800 \over M_{\Sigma} \rm{(GeV)}}} $, in dependence on heavy 5-plet mass $M_{\Sigma}$.}
\label{fig:decay020}
\end{figure}

Let us start our list of the partial decay widths by the decays of neutral $\Sigma^0$ state:
\begin{eqnarray}
&&\Gamma(\Sigma^0\to \ell^-W^+)= {g^2\over 32\pi}\Big|{3\over \sqrt2} V_1^{\ell \Sigma}\Big|^2 {M_\Sigma^3\over M_W^2}\left(1-{M_W^2\over
M_\Sigma^2}\right)^2\left(1+2{M_W^2\over M_\Sigma^2}\right),\nonumber\\
&&\Gamma(\Sigma^0\to \ell^+W^-)= {g^2\over 32\pi}\Big|{\sqrt 3\over 2} V_2^{\ell \Sigma}\Big|^2 {M_\Sigma^3\over M_W^2}\left(1-{M_W^2\over
M_\Sigma^2}\right)^2\left(1+2{M_W^2\over M_\Sigma^2}\right),\nonumber\\
&&\sum^3_{m=1}\Gamma(\Sigma^0\to \nu_m Z^0)= {g^2\over 32\pi c_W^2} \sum_{\ell=\{e, \mu, \tau \}} \Big( \Big|{3\over 2} V_1^{\ell \Sigma}\Big|^2 + \Big|{\sqrt 3\over 2 \sqrt2} V_2^{\ell \Sigma}\Big|^2\Big)\nonumber\\
&& {M_\Sigma^3\over M_Z^2}\left(1-{M_Z^2\over
M_\Sigma^2}\right)^2\left(1+2{M_Z^2\over M_\Sigma^2}\right).
\end{eqnarray}
These three decays, for two chosen set of couplings $V_1$ and $V_2$, are displayed  on Figs.~\ref{fig:decay0800}
and \ref{fig:decay020} as a counterpart of Fig.~2 in Ref.~\cite{Franceschini:2008pz} for type-III triplets. Thereby, the product of $V_1$ and $V_2$ is constrained by Eq.~(\ref{effseesaw}). Decays of $\Sigma^0$ to leptons (antileptons) are governed by $V_1$ ($V_2$), so that the branching ratios to leptons or antileptons are strongly dependent on the hierarchy between $V_1$ and $V_2$, as illustrated in Figs.~\ref{fig:decay0800} and \ref{fig:decay020}.

Besides the representative $\Sigma^0$ decays above,  let us present also the rest of the pointlike decays.
For a negative singly-charged heavy lepton $\Sigma^-$ the partial decay widths are given by:
\begin{eqnarray}
&&\Gamma(\Sigma^-\to \ell^-Z^0)= {g^2\over 32\pi c_W^2} \Big| 3 V_1^{\ell \Sigma}\Big|^2 {M_\Sigma^3\over M_Z^2}\left(1-{M_Z^2\over
M_\Sigma^2}\right)^2\left(1+2{M_Z^2\over M_\Sigma^2}\right),\nonumber\\
&&\sum^3_{m=1}\Gamma(\Sigma^-\to \nu_m W^-)= {g^2\over 32\pi} \sum_{\ell=\{e, \mu, \tau \}} \Big| \sqrt3 V_2^{\ell \Sigma}\Big|^2 \nonumber\\
&&{M_\Sigma^3\over M_W^2}\left(1-{M_W^2\over M_\Sigma^2}\right)^2\left(1+2{M_W^2\over M_\Sigma^2}\right).
\end{eqnarray}
The positive singly-charged $\Sigma^+$ state has the following partial decay widths:
\begin{eqnarray}
&&\Gamma(\Sigma^+\to \ell^+Z^0)= {g^2\over 32\pi c_W^2}\Big|{1\over 2} V_2^{\ell \Sigma}\Big|^2 {M_\Sigma^3\over M_Z^2}\left(1-{M_Z^2\over
M_\Sigma^2}\right)^2\left(1+2{M_Z^2\over M_\Sigma^2}\right),\nonumber\\
&&\sum^3_{m=1}\Gamma(\Sigma^+\to \nu_m W^+)= {g^2\over 32\pi} \sum_{\ell=\{e, \mu, \tau \}} \Big( \Big| \sqrt 3 V_1^{\ell \Sigma}\Big|^2 + \Big| \sqrt2 V_2^{\ell \Sigma}\Big|^2\Big) \nonumber\\
&&{M_\Sigma^3\over M_W^2}\left(1-{M_W^2\over M_\Sigma^2}\right)^2\left(1+2{M_W^2\over M_\Sigma^2}\right).
\end{eqnarray}
Finally, the doubly-charged $\Sigma^{++}$ state decays exclusively via a charged current, with the partial decay width
\begin{equation}\label{decaysigma++}
    \Gamma(\Sigma^{++}\to \ell^+W^+)= {g^2\over 32\pi} \Big| \sqrt3 V_2^{\ell \Sigma}\Big|^2 {M_\Sigma^3\over M_W^2}\left(1-{M_W^2\over
M_\Sigma^2}\right)^2\left(1+2{M_W^2\over M_\Sigma^2}\right).
\end{equation}
Let us stress that there is no such pointlike decay for the triply-charged $\Sigma^{+++}$. This state, instead, has some other interesting decays studied in the next subsections.

\begin{table}[h]
\centering
\begin{tabular}{|c||c|c|c|c|} \hline
                         &  $\overline{\Sigma^+}\to \ell^-Z^0$   &   $\overline{\Sigma^0}\to \ell^+W^-$   &   $\overline{\Sigma^0}\to \ell^-W^+$   &  $\overline{\Sigma^-}\to \ell^+Z^0$ \\\hline
\hline
$\Sigma^+\to \ell^+Z^0$   &  $\ell^+ \ell^-Z^0 Z^0 $   &   $\ell^+ \ell^+Z^0 W^- $     &   $\ell^+ \ell^-Z^0 W^+ $     &  -  \\\hline
$\Sigma^0\to \ell^-W^+$   &  $\ell^- \ell^-W^+ Z^0 $   &   $\ell^- \ell^+W^+ W^- $     &   $\ell^- \ell^-W^+ W^+ $     &  $\ell^- \ell^+W^+ Z^0 $ \\\hline
$\Sigma^0\to \ell^+W^-$   &  $\ell^+ \ell^-W^- Z^0 $   &   $\ell^+ \ell^+W^- W^- $     &   $\ell^+ \ell^-W^- W^+ $     &  $\ell^+ \ell^+W^- Z^0 $ \\\hline
$\Sigma^-\to \ell^-Z^0$   &  -    &   $\ell^- \ell^+Z^0 W^- $     &   $\ell^- \ell^-Z^0 W^+ $     &  $\ell^- \ell^+Z^0 Z^0 $ \\\hline
\end{tabular}
\caption{\footnotesize Decays of exotic leptons to SM particles including same sign dilepton events} 
\label{events}
\end{table}
In Table~\ref{events} we list all possible events coming from the decays of the neutral and singly-charged 5-plet states to the SM charged leptons. This includes the same-sign dilepton events as a distinguished signature at the LHC.

\subsection{Decays which do not depend on any free parameter}

Besides previous decays, there are also decays of $\Sigma^i$ to a lighter $\Sigma^j$ state. The decay rate for a single pion finale state is given by
\begin{equation}\label{widthpion}
\Gamma(\Sigma^i \to \Sigma^j \pi^+ )=  (g^{W\Sigma})^2_{ij} {2 \over \pi}  G_{\rm F}^2 |V_{ud}|^2 f_\pi^2 (\Delta M_{ij})^3
\sqrt{1-\frac{m_\pi^2}{(\Delta M_{ij})^2}}\ .
\end{equation}
The corresponding leptonic decay is
\begin{equation}\label{widthlepton}
\Gamma(\Sigma^i \to \Sigma^j l^+\nu  ) =
(g^{W\Sigma})^2_{ij}  {2 \over 15 \pi^3} G_{\rm F}^2 (\Delta M_{ij})^5 \sqrt{1-\frac{m_l^2}{(\Delta M_{ij})^2}} 
\end{equation}
where $(g^{W\Sigma})^2_{ij}$ is given in Eq.~(\ref{vectorcouplings}) and $\Delta M_{ij}=M_i-M_j$ is given in Eq.~(\ref{massdifferences}).

Such decays are suppressed by small mass differences, except for $\Sigma^{+++}$ where the mass difference $\Delta M_{32}$ between the $\Sigma^{+++}$ and $\Sigma^{++}$ state is large enough to enable also the two pions in the  final state. This three-body decay is dominated  by a $\rho(770)$ resonance, and we obtain for it an analytical formula expressed in terms of the meson decay constants, $f_{\pi}\simeq$  130 MeV and $f_{\rho}\simeq$ 150 MeV,
\begin{eqnarray}\label{widthrho}
\Gamma(\Sigma^{+++} \to \Sigma^{++}  \pi^+  \pi^0)   \simeq \Gamma(\Sigma^{+++} \to \Sigma^{++}  \rho^+ ) = \nonumber \\[3mm]
{24 \over \pi} G_{\rm F}^2 V_{ud}^2 f_{\rho}^2 (\Delta M_{32})^3 \Big(1- \frac{4 m_{\pi}^2}{m_{\rho}^2}\Big)^{-1}
\sqrt{1-\frac{m_{\rho}^2}{(\Delta M_{32})^2}}\ .
\end{eqnarray}
This decay shows an enhancement with respect to the corresponding two-body decay
\begin{equation}
{\Gamma(\Sigma^{+++} \to \Sigma^{++}  \pi^+  \pi^0) \over \Gamma(\Sigma^{+++} \to \Sigma^{++} \pi^+ )} = 6 {f_{\rho}^2 \over f_{\pi}^2}  {\sqrt{1-\frac{m_{\rho}^2}{(\Delta M_{32})^2}} \over \sqrt{1-\frac{m_{\pi}^2}{(\Delta M_{32})^2}}} \Big(1- \frac{4 m_{\pi}^2}{m_{\rho}^2}\Big)^{-1}\ .
\end{equation}
These decays will serve as the referent decays for the golden decay mode of the triply-charged state
studied in the following subsection.

\subsection{Golden $\Sigma^{+++} \to W^+ W^+ l^+$ decay}

We now turn to evaluation of the decay rate for $\Sigma^{+++} \to W^+ W^+ l^+$ with on-shell $W$ bosons. 
\begin{figure}[h]
\begin{center}
\includegraphics[scale=0.9]{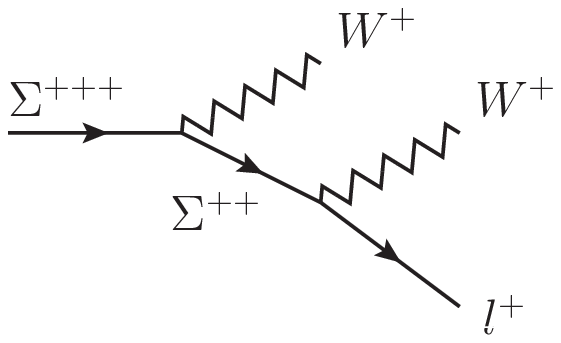}
\end{center}
\caption{Feynman diagram contributing to the triply-charged golden decay}
\label{fig:golden-decay}
\end{figure}
The Feynman diagram on Fig.~\ref{fig:golden-decay} and its crossed, contribute with an off-shell $\Sigma^{++}$ in the intermediate state.

The expression for the amplitude squared is lengthy so we give the analytical expression for decay width only in the limit $M_\Sigma \gg M_W$:
\begin{equation}\label{eq:SPPPwidthlimit}
 \Gamma(\Sigma^{+++} \to W^+ W^+ l^+)\big|_{M_\Sigma\gg M_W}
= {g^2\over 384\pi^2 s_W^2} \Big| \sqrt3 V_2^{\ell \Sigma}\Big|^2 {M_\Sigma^5\over M_W^4}\ .
\end{equation}

\begin{figure}[h]
\begin{center}
\includegraphics[scale=0.7]{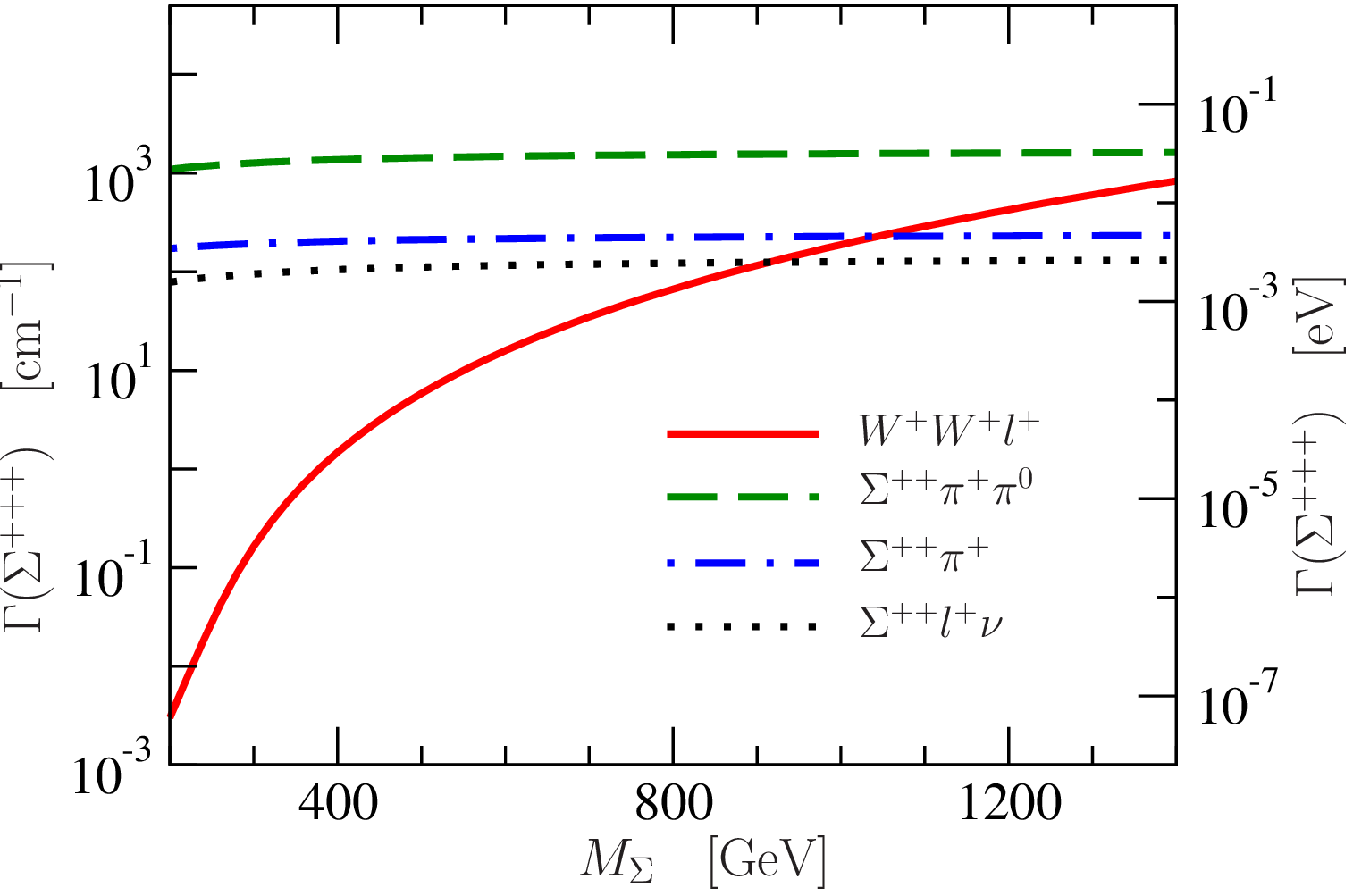}
\end{center}
\caption{Selected partial decay widths of $\Sigma^{+++}$
5-plet lepton for $|V^{l\Sigma}_2| = 10^{-6} \sqrt{{20 \over M_{\Sigma} \rm{(GeV)}}} $ in dependence
of heavy 5-plet mass $M_{\Sigma}$.}
\label{fig:SPPPdecay20}
\end{figure}

\begin{figure}[h]
\begin{center}
\includegraphics[scale=0.7]{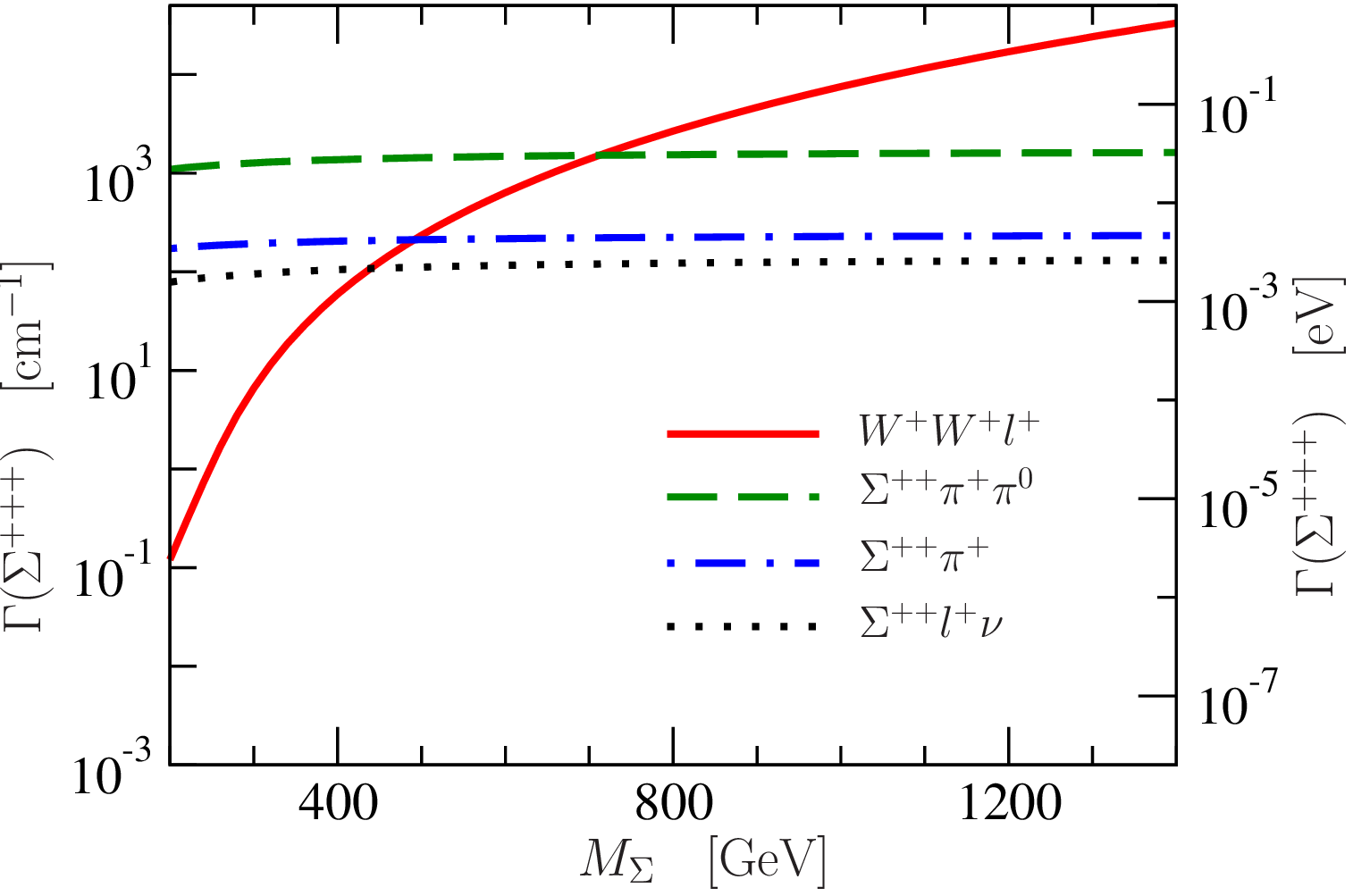}
\end{center}
\caption{Selected partial decay widths of $\Sigma^{+++}$
5-plet lepton for $|V^{l\Sigma}_2| = 10^{-6} \sqrt{{800 \over M_{\Sigma} \rm{(GeV)}}}$ in dependence
of heavy 5-plet mass $M_{\Sigma}$.}
\label{fig:SPPPdecay800}
\end{figure}

Obviously, this decay is governed by the same mixing factor as in $\Sigma^{++}$ decay in Eq.~(\ref{decaysigma++}). Full numerical calculation, using FormCalc
package \cite{Hahn:1998yk}, results in a partial width plotted on Figs.~\ref{fig:SPPPdecay20} and \ref{fig:SPPPdecay800} for same choices of the mixing factor $V_2$ as on previous figures. On the same figures we plot the partial widths of the decays of $\Sigma^{+++}$ given in Eqs.~(\ref{widthpion})-(\ref{widthrho}).

For the sake of being definite, let us explicate the branching ratios for the choice of heavy-light mixing $V_2$ taken on Fig.~\ref{fig:SPPPdecay800}, for two values of heavy-lepton mass $M_\Sigma$
\begin{eqnarray}
  \sum_l BR(\Sigma^{+++} \to W^+ W^+ l^+))\Big|_{M_\Sigma = 400 \rm{GeV}} &=& 0.09 \nonumber\\
  \sum_l BR(\Sigma^{+++} \to W^+ W^+ l^+))\Big|_{M_\Sigma = 800 \rm{GeV}} &=& 0.80 \ .
\end{eqnarray}

As stated before, for $1\ \rm{fb^{-1}}$ of integrated luminosity expected to be collected from the present LHC run at $\sqrt{s}=7$ TeV, there will be $\sim 100$ triply-charged $\Sigma^{+++}$ or $\overline{\Sigma^{+++}}$ fermions produced. This will result in $\sim$10 decays $\Sigma^{+++}(\overline{\Sigma^{+++}}) \to W^{\pm} W^{\pm} l^{\pm}$ for $M_\Sigma = 400\ \rm{GeV}$ and $|V^{l\Sigma}_2| = 10^{-6} \sqrt{{800\ \over M_{\Sigma} \rm{(GeV)}}}$.

\section{Conclusions}
\label{sec:concl}

We have investigated the testability of exotic heavy leptons introduced to explain small neutrino masses. It is conceivable that such states additional to the SM leptons may appear as higher representations such as isospin 2 heavy fermions in conjunction with
isospin 3/2 heavy scalar fields studied here. The LHC could provide enough energy and luminosity to produce and study such heavy
particles if they mediate the proposed seesaw mechanism. In particular, the decays of new heavy leptons to SM particles depend on how light and heavy leptons mix with each other. Our seesaw model constrains the mixing to be small, given by the square root of the ratio of
light and small masses, $(m_{\nu}/M_{\Sigma})^{1/2}$. This suppresses their effects on the flavor physics, but the abundant production of new states invites us to investigate the phenomenology of these heavy leptons at the LHC.

For $1\ \rm{fb^{-1}}$ of integrated luminosity, the present LHC ($\sqrt{s}=7$ TeV) run can produce in total 270 (7000) $\Sigma$-$\overline \Sigma$ pairs of heavy leptons with mass $M_\Sigma=\rm{400\ GeV\ (200\ GeV)}$, among which there would be $\sim 100\ (2850)$ triply-charged $\Sigma^{+++}$ or $\overline{\Sigma^{+++}}$ fermions. Thereby, the triply-charged $\Sigma^{+++}$ has the biggest production cross-section, $\sigma(\Sigma^{+++})|_{M_\Sigma = 400 \rm{GeV}}= 63.1\ \rm{fb}$, and singly-charged $\Sigma^{-}$ has the smallest but still comparable $\sigma(\Sigma^{-})|_{M_\Sigma = 400 \rm{GeV}}= 43.7\ \rm{fb}$. The production cross-sections for all other particles and antiparticles from a 5-plet are in between. Notably, the production of triply-charged leptons is followed with its
golden decays $\Sigma^{+++}(\overline{\Sigma^{+++}})\to W^{\pm} W^{\pm} l^{\pm}$. For a chosen mass $M_\Sigma = 400\ \rm{GeV}$ and the mixing $|V^{l\Sigma}_2| = 10^{-6} \sqrt{{800\ \over M_{\Sigma} \rm{(GeV)}}}$ we expect $\sim$10 golden mode events. 

The great advantage of the processes considered in this paper,
common to most seesaw-inspired models, is the
copious production of energetic leptons, which provides a clear
experimental signature and relatively small SM background.
Detailed studies of standard seesaw scenarios \cite{delAguila:2008cj} demonstrate that
final states with 2 (same sign) and 3 or 4 charged leptons
can be increasingly easily distinguished from the SM background.
For the  model at hand, similar results are expected. 
For example, our prominent production channel $p \bar{p} \to \Sigma^{+++} \overline{\Sigma^{+++}}$
results in 3 or more energetic charged leptons in roughly one third of cases.
Additionally, when it is experimentally possible to separate final-state particles
coming from each of the two produced heavy leptons, then the three same-sign
leptons coming from a single heavy-lepton lead to distinctive 
signature of the presented model. The corresponding SM background is essentially negligible, similarly to what has been demonstrated in \cite{Mukhopadhyay:2011xs} in generic new physics scenario with lepton number violation.

To conclude, the exotic 5-plet leptons introduced to explain small neutrino masses can be observed at LHC. At the designed LHC luminosity and $\sqrt{s}=14$ TeV our 5-plet tree level seesaw model is falsifiable for sub-TeV $\Sigma$-lepton masses, while for masses which are outside of the LHC reach the 5-plet leptons remain viable for generating radiative neutrino masses. In this sense, we hope that the involved predictive and testable vectorlike fermion 5-plet may establish an exciting link between collider phenomenology and the origin of neutrino masses.\\

\subsubsection*{Acknowledgements}

We thank  Borut Bajc, Walter Grimus and Goran Senjanovi\'c for discussions at the early stage of this research. K.K. and I.P. thank Jan O. Eeg for discussions and hospitality offered at the University in Oslo. This work is supported by the Croatian Ministry  of Science, Education, and Sports under Contract No. 119-0982930-1016.

\subsection*{APPENDIX: The interaction vertices of the 5-plet}

The novel seesaw model employs, in addition to the SM particles, $n_\Sigma$
vectorlike 5-plets of leptons with hypercharge two,
$\Sigma_{L,R} \sim (1,5,2)$ under $SU(3)_C\times SU(2)_L\times
U(1)_Y$. We write the component fields as
\begin{eqnarray}
&&\Sigma_L=
\left(
  \begin{array}{c}
    \Sigma_L^{+++}\\
    \Sigma_L^{++}\\
    \Sigma_L^{+}\\
    \Sigma_L^{0}\\
    \Sigma_L^{-}\\
  \end{array}
\right)\ \ \ ,\ \ \
\Sigma_R=
\left(
  \begin{array}{c}
    \Sigma_R^{+++}\\
    \Sigma_R^{++}\\
    \Sigma_R^{+}\\
    \Sigma_R^{0}\\
    \Sigma_R^{-}\\
  \end{array}
\right)\ .
\end{eqnarray}
The renormalizable Lagrangian involving $\Sigma_L$ and $\Sigma_R$ is given by
\begin{eqnarray}
\mathcal{L}&=& \overline{\Sigma_L}i\cancel{D}\Sigma_L + \overline{\Sigma_R}i\cancel{D}\Sigma_R
-\overline{\Sigma_R}M_\Sigma \Sigma_L -\overline{\Sigma_L}M_\Sigma^\dagger \Sigma_R \nonumber \\
&+&\overline{\Sigma_R}Y_1L_L\Phi_1^* + \overline{(\Sigma_L)^c} Y_2 L_L \Phi_2+ \mathrm{H.c.}\ .
\end{eqnarray}
With a nonzero vacuum expectation value for the scalar fields, doublet leptons receive mass and  mix with the 5-plet leptons.
Lepton mass terms in the Lagrangian are
\begin{eqnarray}
\mathcal{L}_m&=&-\frac{1}{2}
\left(  \overline{(\nu_L)^c}  \; \overline{(\Sigma_L^0)^c} \; \overline{\Sigma_R^0} \right)
\left( \! \begin{array}{ccc}
0 & m_2^T & m_1^T \\
m_2 & 0 & M_{\Sigma}^T \\
m_1 & M_{\Sigma} & 0
\end{array} \! \right) \,
\left( \!\! \begin{array}{c} \nu_L \\ \Sigma_L^0 \\ (\Sigma_R^0)^c \end{array} \!\! \right)
\; + \mathrm{H.c.}\nonumber\\
&-&\left( \overline{l_R} \; \overline{\Sigma_R^-} \; \overline{(\Sigma_L^+)^c} \right)
\left( \! \begin{array}{ccc}
m_l & 0 & 0 \\
m_3 & M_{\Sigma} & 0 \\
m_4 & 0 & M_{\Sigma}^T
\end{array} \! \right) \,
\left( \!\! \begin{array}{c} l_L \\ \Sigma_L^- \\ (\Sigma_R^+)^c \end{array} \!\! \right)
\; + \mathrm{H.c.}\nonumber\\
&-&
\overline{\Sigma_R^{++}} M_\Sigma  \Sigma_L^{++} -  \overline{\Sigma_R^{+++}} M_\Sigma \Sigma_L^{+++} + \mathrm{H.c.}\ ,
\label{mass-matrix}
\end{eqnarray}
with
\begin{eqnarray}
m_1 = \sqrt{\frac{1}{10}}Y_1{v^*_\Phi}_1\ , && m_2 = - \sqrt{\frac{3}{20}} Y_2{v_\Phi}_2\ , \nonumber \\
m_3 = \sqrt{\frac{2}{5}}Y_1{v^*_\Phi}_1\ , && m_4 = \sqrt{\frac{1}{10}} Y_2{v_\Phi}_2\ .
\end{eqnarray}
For detailed studies, one needs to understand the mass matrices in Eq.~(\ref{mass-matrix}) and their diagonalization.
The diagonalization of the mass matrices can be achieved by making the following unitary transformations on the leptons
\begin{equation}\nonumber
    \left(
    \begin{array}{c}
      \nu_{L} \\
      \Sigma^0_{L} \\
      (\Sigma_R^0)^c \\
    \end{array}
  \right)
  =U^0
  \left(
    \begin{array}{c}
      \nu_{mL} \\
      \Sigma^0_{mL} \\
      (\Sigma_{mR}^0)^c \\
    \end{array}
  \right)\ ,
\end{equation}
\begin{equation}
    \left(
    \begin{array}{c}
      l_L \\
      \Sigma_L^- \\
      (\Sigma_R^+)^c \\
    \end{array}
  \right)
  =U^L
  \left(
    \begin{array}{c}
      l_{mL} \\
      \Sigma_{mL}^- \\
      (\Sigma_{mR}^+)^c \\
    \end{array}
  \right)\ , \ \
\left(
    \begin{array}{c}
      l_R \\
      \Sigma_R^- \\
      (\Sigma_L^+)^c \\
    \end{array}
  \right)
  =U^R
  \left(
    \begin{array}{c}
      l_{mR} \\
      \Sigma_{mR}^- \\
      (\Sigma_{mL}^+)^c \\
    \end{array}
  \right)\ ,
\end{equation}
different for a symmetric neutral  and a nonsymmetric charged mass matrix
\begin{eqnarray}
  U^{0^T}\left( \! \begin{array}{ccc}
0 & m_2^T & m_1^T \\
m_2 & 0 & M_{\Sigma}^T \\
m_1 & M_{\Sigma} & 0
\end{array} \! \right)
U^0 &=& \left( \! \begin{array}{cc}
m_\nu & 0 \\
0 & M_0
\end{array} \! \right)\ , \nonumber\\
  U^{R^\dagger} \left( \! \begin{array}{ccc}
m_l & 0 & 0 \\
m_3 & M_{\Sigma} & 0 \\
m_4 & 0 & M_{\Sigma}^T
\end{array} \! \right) U^L &=& \left( \! \begin{array}{cc}
m_{lepton} & 0 \\
0 & M
\end{array} \! \right)\ .
\end{eqnarray}
For 3 light doublet fields and $n_\Sigma$ heavy 5-plet vectorlike lepton fields the matrices $U_{L,R}$ and $U_0$ are $(3+2n_\Sigma)\times(3+2n_\Sigma)$ unitary matrices, which we decompose into the block matrices as follows:
\begin{equation}
U^0\equiv \left(
  \begin{array}{ccc}
    U^0_{\nu\nu} & U^0_{\nu \Sigma_L^0} & U^0_{\nu(\Sigma_R^0)^c}\\
    U^0_{\Sigma_L^0\nu} & U^0_{\Sigma_L^0\Sigma_L^0} & U^0_{\Sigma_L^0(\Sigma_R^0)^c}\\
    U^0_{(\Sigma_R^0)^c\nu} & U^0_{(\Sigma_R^0)^c\Sigma_L^0} & U^0_{(\Sigma_R^0)^c(\Sigma_R^0)^c}\\
  \end{array}
\right)\ ,
\end{equation}
\begin{equation}
U^L\equiv \left(
  \begin{array}{ccc}
    U^L_{ll} & U^L_{l\Sigma_L^-} & U^L_{l(\Sigma_R^+)^c}\\
    U^L_{\Sigma_L^-l} & U^L_{\Sigma_L^-\Sigma_L^-} & U^L_{\Sigma_L^-(\Sigma_R^+)^c}\\
    U^L_{(\Sigma_R^+)^cl} & U^L_{(\Sigma_R^+)^c\Sigma_L^-} & U^L_{(\Sigma_R^+)^c(\Sigma_R^+)^c}\\
  \end{array}
\right)\ ,
\end{equation}
\begin{equation}
U^R\equiv \left(
  \begin{array}{ccc}
    U^R_{ll} & U^R_{l\Sigma_R^-} & U^R_{l(\Sigma_L^+)^c}\\
    U^R_{\Sigma_R^-l} & U^R_{\Sigma_R^-\Sigma_R^-} & U^R_{\Sigma_R^-(\Sigma_L^+)^c}\\
    U^R_{(\Sigma_L^+)^cl} & U^R_{(\Sigma_L^+)^c\Sigma_R^-} & U^R_{(\Sigma_L^+)^c(\Sigma_L^+)^c}\\
  \end{array}
\right)\ .
\end{equation}
Here, we distinguish the entries $U^0_{\nu\nu}$, $U^L_{ll}$ and $U^R_{ll}$ which are $3\times3$ matrices, the entries $U^0_{\nu \Sigma_L^0}$, $U^0_{\nu(\Sigma_R^0)^c}$, $U^L_{l\Sigma_L^-}$, $U^L_{l(\Sigma_R^+)^c}$, $U^R_{l\Sigma_R^-}$ and $U^R_{l(\Sigma_L^+)^c}$ which are $3\times n_\Sigma$ matrices, and the entries $U^0_{\Sigma_L^0\nu}$, $U^0_{(\Sigma_R^0)^c\nu}$, $U^L_{\Sigma_L^-l}$, $U^L_{(\Sigma_R^+)^cl}$, $U^R_{\Sigma_R^-l}$ and $U^R_{(\Sigma_L^+)^cl}$ which are $n_\Sigma\times3$ matrices. The remaining entries are $n_\Sigma \times n_\Sigma$ matrices.

In principle, the matrices $U_{0}$ and $U_{L,R}$ can be expressed in terms of $m_{1,2,3,4}$, $m_l$ and $M_\Sigma$ sub-blocks. Since, in order for the seesaw mechanism to work, the factors $m_{1,2,3,4} M^{-1}_\Sigma$ should be small, one can expand $U_{0}$ and $U_{L,R}$ in powers of $M^{-1}_\Sigma$ and keep a track of the leading order contributions. For this purpose, without loss of generality, it is convenient to write the leading order expressions up to $M^{-2}_\Sigma$ in the basis where $m_l$ and $M_\Sigma$ are already diagonalized.
Following the procedure of Refs.~\cite{Li:2009mw, Grimus:2000vj}  we first obtain results for the entries $U^0$:
\begin{eqnarray}
&&U^0_{\nu\nu} = (1 - \frac{1}{2} m_1^\dagger M^{-2}_\Sigma m_1 - \frac{1}{2} m_2^\dagger M^{-2}_\Sigma m_2)V_{PMNS}, \ \
U^0_{\nu \Sigma_L^0} = m_1^\dagger M^{-1}_\Sigma, \ \ \nonumber\\
&&U^0_{\nu(\Sigma_R^0)^c} = m_2^\dagger M^{-1}_\Sigma, \ \
U^0_{\Sigma_L^0\nu} = -M^{-1}_\Sigma m_1 V_{PMNS}, \ \ \nonumber\\
&&U^0_{\Sigma_L^0\Sigma_L^0} = 1 - \frac{1}{2} M^{-1}_\Sigma m_1 m_1^\dagger M^{-1}_\Sigma, \ \
U^0_{\Sigma_L^0(\Sigma_R^0)^c} = - \frac{1}{2} M^{-1}_\Sigma m_1 m_2^\dagger M^{-1}_\Sigma, \ \ \nonumber\\
&&U^0_{(\Sigma_R^0)^c\nu} = -M^{-1}_\Sigma m_2 V_{PMNS}, \ \
U^0_{(\Sigma_R^0)^c\Sigma_L^0} = - \frac{1}{2} M^{-1}_\Sigma m_2 m_1^\dagger M^{-1}_\Sigma, \ \ \nonumber\\
&&U^0_{(\Sigma_R^0)^c(\Sigma_R^0)^c} = 1 - \frac{1}{2} M^{-1}_\Sigma m_2 m_2^\dagger M^{-1}_\Sigma\ . \label{U0razvoj}
\end{eqnarray}
Here, $V_{PMNS}$ is a $3 \times 3$ unitary matrix which diagonalizes the effective light neutrino mass matrix:
\begin{equation}\label{effective}
    \tilde{m}_\nu\simeq - m_2^T M^{-1}_\Sigma m_1 - m_1^T M^{-1}_\Sigma m_2\ ,\ \ V_{PMNS}^T \tilde{m}_\nu V_{PMNS} = m_\nu \ .
\end{equation}
The mass matrix for  heavy neutral leptons acquires corrections of the order $M^{-1}_\Sigma$,
\begin{eqnarray}\label{neutralcorr}
\nonumber
\tilde{M}_0  \simeq \left( \begin{array}{cc}
0 & M_{\Sigma} \\
M_{\Sigma} & 0
\end{array} \right) &+& \frac{1}{2} \left( \begin{array}{cc}
m_2 m_1^\dagger M^{-1}_\Sigma & m_2 m_2^\dagger M^{-1}_\Sigma \\
m_1 m_1^\dagger M^{-1}_\Sigma & m_1 m_2^\dagger M^{-1}_\Sigma
\end{array} \right) \\
&+& \frac{1}{2} \left( \begin{array}{cc}
M^{-1}_\Sigma m_1^* m_2^T  & M^{-1}_\Sigma m_1^* m_1^T \\
M^{-1}_\Sigma m_2^* m_2^T  & M^{-1}_\Sigma m_2^* m_1^T
\end{array} \right) \; ,
\end{eqnarray}
which, being of the order of light neutrino masses, are completely negligible for our phenomenological considerations.
In principle, for every vectorlike 5-plet we would get two nearly degenerate heavy Majorana fermions, with mass splitting of the order $m_\nu$. By neglecting these corrections we treat heavy neutral leptons as Dirac fermions for all practical phenomenological considerations.\\

Next, we obtain the leading order expressions up to $M^{-2}_\Sigma$  for the entries in $U^L$:
\begin{eqnarray}
&&U^L_{ll} = 1 - \frac{1}{2} m_3^\dagger M^{-2}_\Sigma m_3 - \frac{1}{2} m_4^\dagger M^{-2}_\Sigma m_4, \ \
U^L_{l\Sigma_L^-} = m_3^\dagger M^{-1}_\Sigma, \ \ \nonumber\\
&&U^L_{l(\Sigma_R^+)^c} = m_4^\dagger M^{-1}_\Sigma, \ \
U^L_{\Sigma_L^-l} = -M^{-1}_\Sigma m_3, \ \ \nonumber\\
&&U^L_{\Sigma_L^-\Sigma_L^-} = 1 - \frac{1}{2} M^{-1}_\Sigma m_3 m_3^\dagger M^{-1}_\Sigma, \ \
U^L_{\Sigma_L^-(\Sigma_R^+)^c} = - \frac{1}{2} M^{-1}_\Sigma m_3 m_4^\dagger M^{-1}_\Sigma, \ \ \nonumber\\
&&U^L_{(\Sigma_R^+)^cl} = -M^{-1}_\Sigma m_4, \ \
U^L_{(\Sigma_R^+)^c\Sigma_L^-} = - \frac{1}{2} M^{-1}_\Sigma m_4 m_3^\dagger M^{-1}_\Sigma, \ \ \nonumber\\
&&U^L_{(\Sigma_R^+)^c(\Sigma_R^+)^c} = 1 - \frac{1}{2} M^{-1}_\Sigma m_4 m_4^\dagger M^{-1}_\Sigma\ . \label{ULrazvoj}
\end{eqnarray}
Similarly, for $U^R$, the leading order expressions up to $M^{-2}_\Sigma$ are:
\begin{eqnarray}
&&U^R_{ll} = 1, \ \
U^R_{l\Sigma_R^-} = m_l m_3^\dagger M^{-2}_\Sigma, \ \
U^R_{l(\Sigma_L^+)^c} = m_l m_4^\dagger M^{-2}_\Sigma, \ \ \nonumber\\
&&U^R_{\Sigma_R^-l} = -M^{-2}_\Sigma m_3 m_l, \ \
U^R_{\Sigma_R^-\Sigma_R^-} = 1, \ \
U^R_{\Sigma_R^-(\Sigma_L^+)^c} = 0, \ \ \nonumber\\
&&U^R_{(\Sigma_L^+)^cl} = -M^{-2}_\Sigma m_4 m_l, \ \
U^R_{(\Sigma_L^+)^c\Sigma_R^-} = 0, \ \
U^R_{(\Sigma_L^+)^c(\Sigma_L^+)^c} = 1. \label{URrazvoj}
\end{eqnarray}
As a result, the mass matrix for the light singly-charged leptons gets corrections of the order $M^{-2}_\Sigma$
\begin{equation} \label{mlrazvoj}
    \tilde{m}_l \simeq m_l (1 - \frac{1}{2} m_3^\dagger M^{-2}_\Sigma m_3 - \frac{1}{2} m_4^\dagger M^{-2}_\Sigma m_4)\ ,
\end{equation}
while, the mass matrix for the heavy singly-charged leptons gets corrections of the order $M^{-1}_\Sigma$
\begin{equation}\label{Mrazvoj}
\tilde{M}  \simeq \left( \begin{array}{cc}
M_{\Sigma} & 0 \\
0 & M_{\Sigma}
\end{array} \right) + \frac{1}{2} \left( \begin{array}{cc}
m_3 m_3^\dagger M^{-1}_\Sigma & m_3 m_4^\dagger M^{-1}_\Sigma \\
m_4 m_3^\dagger M^{-1}_\Sigma & m_4 m_4^\dagger M^{-1}_\Sigma
\end{array} \right)\ .
\end{equation}
Both corrections, in Eq.~(\ref{mlrazvoj}) and Eq.~(\ref{Mrazvoj}), are completely negligible for our phenomenological considerations.

Our studies of particle  production require a knowledge of gauge couplings to leptonic fields. In the weak interaction basis, they can be written
as
\begin{eqnarray}\label{gauge-complete}
\mathcal{L}_{gauge}=
&+& e (3\overline{\Sigma^{+++}}\gamma^\mu \Sigma^{+++} + 2\overline{\Sigma^{++}}\gamma^\mu \Sigma^{++}
+ \overline{\Sigma^{+}}\gamma^\mu \Sigma^{+} - \overline{\Sigma^{-}}\gamma^\mu \Sigma^{-}
-\overline{l }\gamma^\mu l)A_\mu \nonumber \\
&+&{g \over c_W}( (2-3s_W^2) \overline{\Sigma^{+++}}\gamma^\mu \Sigma^{+++} + (1-2s_W^2) \overline{\Sigma^{++}}\gamma^\mu\Sigma^{++})Z_\mu\nonumber\\
&-& {g \over c_W}s_W^2 (\overline{\Sigma^{+}}\gamma^\mu \Sigma^{+} - \overline{\Sigma^{-}}\gamma^\mu \Sigma^{-}
-\overline{l }\gamma^\mu l) Z_\mu\nonumber \\
&+& {g \over c_W} (\bar \nu_L \gamma^\mu \nu_L +
\overline{\Sigma_L^{0}}\gamma^\mu \Sigma_L^{0} - \overline{\Sigma_R^{0}}\gamma^\mu \Sigma_R^{0}) Z_\mu\nonumber\\
&+& {g \over c_W} (-{1\over 2} \bar \nu_L \gamma^\mu \nu_L -2 \overline{\Sigma_L^{0}}\gamma^\mu \Sigma_L^{0}
- {1\over 2} \bar l_L \gamma^\mu l_L
-2 \overline{\Sigma_L^{-}}\gamma^\mu \Sigma_L^{-} - 2 \overline{\Sigma_R^{-}}\gamma^\mu \Sigma_R^{-}) Z_\mu\nonumber\\
&+&g\Big[\sqrt{2}\overline{\Sigma^{++}_L}\gamma^\mu\Sigma^{+++}_L + \sqrt{3}\overline{\Sigma^{+}_L}\gamma^\mu\Sigma^{++}_L
+ \sqrt{3}\overline{\Sigma^{0}_L}\gamma^\mu\Sigma^{+}_L + \sqrt{2}\overline{\Sigma^{-}_L}\gamma^\mu\Sigma^{0}_L  \nonumber \\
&+&\sqrt{2}\overline{\Sigma^{++}_R}\gamma^\mu\Sigma^{+++}_R + \sqrt{3}\overline{\Sigma^{+}_R}\gamma^\mu\Sigma^{++}_R
+ \sqrt{3}\overline{\Sigma^{0}_R}\gamma^\mu\Sigma^{+}_R  + \sqrt{2}\overline{\Sigma^{-}_R}\gamma^\mu\Sigma^{0}_R  \nonumber \\
&+&{1\over\sqrt{2}}\overline{l_L}\gamma^\mu \nu_L \Big] W_\mu^- + \mathrm{H.c.} \;,
\end{eqnarray}
where $c_W = \cos\theta_W$ and $s_W = \sin\theta_W$.

In the mass-eigenstate basis we obtain the terms involving the heavy fermion, light fermion and  gauge boson fields that are relevant for the decays of heavy leptons. In the following, we focus only on the above terms and restrict to couplings of the order $M^{-1}_\Sigma$. Whereas the  photon couplings to all leptons and $Z$ couplings to triply and doubly charged leptons are diagonal in the mass-eigenstate basis, the
couplings of $Z$-boson  to singly-charged and neutral leptons are more complicated. They are given by
\begin{equation}
    \mathcal{L}_{NCZ} \equiv (\mathcal{L}_{NCZ}^A + \mathcal{L}_{NCZ}^B + \mathcal{L}_{NCZ}^C+ \mathrm{H.c.}) \; ,
\end{equation}
where
\begin{eqnarray}
\mathcal{L}_{NCZ}^A&=&{g\over c_W}\overline{l_m}V_{Zl\Sigma^-}^L\gamma^\mu P_L
\Sigma^-_{m'}Z_\mu^0,\nonumber\\
\mathcal{L}_{NCZ}^B&=&{g\over c_W}\overline{(l_m)^c}V_{Zl\Sigma^+}^R\gamma^\mu P_R
\Sigma^+_{m'} Z_\mu^0,\nonumber\\
\mathcal{L}_{NCZ}^C&=&{g\over c_W}[\overline{\nu_m}V_{Z\nu\Sigma^0}^L\gamma^\mu P_L
\Sigma^0_{m'}Z_\mu^0+\overline{\nu_m}V_{Z\nu\Sigma^0}^R\gamma^\mu P_R
\Sigma^0_{m'}Z_\mu^0] \; .\nonumber\\
\end{eqnarray}
Here, the matrix couplings, up to order $M^{-1}_\Sigma$, are given by
\begin{eqnarray}
&&V_{Zl\Sigma^-}^L= -2 U^{L^\dagger}_{\Sigma_L^-l} - \frac{1}{2} U^L_{l\Sigma_L^-} ,\ \
V_{Zl\Sigma^+}^R= \frac{1}{2} U^{L^*}_{l(\Sigma_R^+)^c},\nonumber\\
&&V_{Z\nu\Sigma^0}^L= -2 U^{0^\dagger}_{\Sigma_L^0\nu } - \frac{1}{2} U^{0^\dagger}_{\nu\nu} U^0_{\nu \Sigma_L^0},\ \
V_{Z\nu\Sigma^0}^R= \frac{1}{2} U^{0^T}_{\nu\nu} U^{0^*}_{\nu(\Sigma_R^0)^c}\ .
\end{eqnarray}

Analogously, the charged current interactions are given by
\begin{eqnarray}
\mathcal{L}_{CC}
&\equiv&
(\mathcal{L}_{CC}^A+\mathcal{L}_{CC}^B+\mathcal{L}_{CC}^C+\mathcal{L}_{CC}^D+\mathcal{L}_{CC}^E + \mathrm{H.c.})\; ,
\end{eqnarray}
where
\begin{eqnarray}
\mathcal{L}_{CC}^A&=& g [\overline{\nu_m} V_{\nu \Sigma^+}^{L} \gamma^\mu P_L
\Sigma_{m'}^+W_\mu^- + \overline{\nu_m} V_{\nu \Sigma^+}^{R}\gamma^\mu
P_R \Sigma_{m'}^{+}W_\mu^-],\nonumber\\
\mathcal{L}_{CC}^B&=& g [\overline{\Sigma^-_m} V^L_{\Sigma^-\nu}\gamma^\mu P_L \nu_{m'}
W^-_\mu + \overline{\Sigma_m} V^R_{\Sigma^-\nu}\gamma^\mu P_R
\nu_{m'}W^-_\mu],\nonumber\\
\mathcal{L}_{CC}^C&=& g \overline{l_m} V_{l \Sigma^0}^{L}\gamma^\mu P_L \Sigma^0_{m'}W_\mu^- ,\nonumber\\
\mathcal{L}_{CC}^D&=& g \overline{\Sigma^0_m} V^R_{\Sigma^0l}\gamma^\mu P_R (l_{m'})^c
W^-_\mu,\nonumber\\
\mathcal{L}_{CC}^E&=& g \overline{(l_m)^c} V_{l \Sigma^{++}}^{R}\gamma^\mu P_R \Sigma^{++}_{m'}W_\mu^- ,
\end{eqnarray}
and up to order $M^{-1}_\Sigma$
\begin{eqnarray}
&&V_{\nu \Sigma^+}^{L}= \sqrt{3} U^{0^\dagger}_{\Sigma_L^0\nu }, \ \
V_{\nu \Sigma^+}^{R}=  \sqrt{3} U^{0^T}_{(\Sigma_R^0)^c \nu} - {1\over\sqrt{2}} U^{0^T}_{\nu\nu} U^{L^*}_{l (\Sigma_R^+)^c},\nonumber\\
&&V^L_{\Sigma^-\nu}= {1\over\sqrt{2}} U^{L^\dagger}_{l \Sigma_L^-} U^0_{\nu\nu} + \sqrt{2} U^0_{\Sigma_L^0 \nu}, \ \
V^R_{\Sigma^-\nu}= \sqrt{2} U^{0^*}_{(\Sigma_R^0)^c \nu},\nonumber\\
&&V_{l \Sigma^0}^{L}= \sqrt{2} U^{L^\dagger}_{\Sigma_L^- l} + {1\over\sqrt{2}} U^0_{\nu \Sigma_L^0},\ \
V^R_{\Sigma^0l}= \sqrt{3} U^{L^*}_{(\Sigma_R^+)^c l} - {1\over\sqrt{2}} U^{0^T}_{\nu (\Sigma_R^0)^c},\nonumber\\
&&V_{l \Sigma^{++}}^{R}= \sqrt{3} U^{L^T}_{(\Sigma_R^+)^c l}.
\end{eqnarray}

Finally, by suppressing the  indices $m$ and $m'$ and by defining
\begin{equation}
    V_1= \sqrt{\frac{1}{10}} Y^\dagger_1{v_\Phi}_1 M^{-1}_\Sigma\ , V_2= \sqrt{\frac{1}{10}} Y^\dagger_2{v^*_\Phi}_2 M^{-1}_\Sigma \ ,
\end{equation}
we obtain more compact expressions

\begin{eqnarray}
  \mathcal{L}_{NCZ} &=& {g\over
c_W}\Big[ \overline{\nu} ( {3\over2} V_{PMNS}^\dagger V_1 \gamma^\mu P_L
+{-\sqrt 3 \over 2 \sqrt 2} V_{PMNS}^T V^*_2 \gamma^\mu P_R ) \Sigma^0 \nonumber \\
&+& \overline{l} (3V_1 \gamma^\mu P_L )\Sigma^- + \overline{l^c} ({1 \over 2} V^*_2  \gamma^\mu P_R )
\Sigma^+ \Big] Z_\mu^0 + \mathrm{H.c.}
\end{eqnarray}
and
\begin{eqnarray}
\mathcal{L}_{CC} &=& g\Big[\overline{\nu}(- \sqrt 3 V_{PMNS}^\dagger V_1 \gamma^\mu P_L + \sqrt 2 V_{PMNS}^T V^*_2 \gamma^\mu P_R)\Sigma^+\nonumber\\
&+& \overline{\Sigma^-} (\sqrt 3 V^T_2 V_{PMNS}^* \gamma^\mu P_R) \nu + \overline{l} (-{3\over \sqrt 2} V_1 \gamma^\mu P_L) \Sigma^0 \nonumber\\
&+& \overline{\Sigma^0} (-{\sqrt 3\over 2} V^T_2 \gamma^\mu P_R) l^c + \overline{l^c} (-\sqrt 3 V^*_2 \gamma^\mu P_R) \Sigma^{++} \Big] W_\mu^- + \mathrm{H.c.}
\end{eqnarray}
Note that the interactions involving light neutrinos in the above have the additional $V_{PMNS}$ factor compared with those involving light charged leptons. For easier comparison to the type-III seesaw case, we are keeping in this Appendix the notation and the style from
Ref.~\cite{Li:2009mw} as much as possible.

\end{document}